\documentclass[prr, reprint, superscriptaddress, amsmath, amssymb, aps]{revtex4-2}

\usepackage{graphicx}
\usepackage{dcolumn}
\usepackage{bm}
\usepackage[protrusion=true,expansion=true]{microtype}
\usepackage[normalem]{ulem}
\usepackage{soul}
\usepackage{xcolor}
\usepackage{kotex}
\usepackage{hyperref}
\usepackage{cleveref}
\usepackage[english]{babel} 

\usepackage{tikz}
\usetikzlibrary{shapes.geometric, positioning, arrows.meta}

\usepackage{adjustbox}

\usepackage{algorithm}
\usepackage{algpseudocode}
\floatname{algorithm}{Algorithm}

\clearpage

\def\max{\textrm{max}}

\begin{document}

\preprint{APS/123-QED}

\title{From Spatial to Spectral: Network Renormalization via Dynamical Correlations}

\author{Cook Hyun Kim}
\affiliation{CCSS, KI for Grid Modernization, Korea Institute of Energy Technology, 21 Kentech-gil, Naju-si, Jeollanam-do, 58330, Republic of Korea}

\author{B. Kahng}
\email{bkahng@kentech.ac.kr}
\affiliation{CCSS, KI for Grid Modernization, Korea Institute of Energy Technology, 21 Kentech-gil, Naju-si, Jeollanam-do, 58330, Republic of Korea}

\date{\today}

\begin{abstract}
Network renormalization has traditionally relied on spatial adjacency—grouping nearby nodes together—but this approach fails to capture the dynamical correlations that govern system-wide behavior in scale-free networks.
We present a spectral-space renormalization framework that enables coarse-graining based on dynamical coherence rather than geometric proximity. Within this framework, diffusion processes naturally constitute renormalization transformations in spectral space, yielding scaling relations that connect network dimensions with critical exponents.
Building on this foundation, we develop a meta-graph reconstruction algorithm that systematically maps spectral information back into explicit topology while preserving dynamical correlations. The resulting renormalized networks uncover organizational structures that remain invisible to adjacency-based methods, including long-range correlations between structurally distant nodes that reflect coherent dynamical responses.
Applications to Internet topologies, yeast regulatory networks, and European power grids demonstrate the broad applicability of this framework. The algorithm consistently extracts fractal ($d_f$), spectral ($d_s$), and random-walk ($d_w$) dimensions with theoretical consistency across diverse systems. In power grids, it further reveals hidden failure pathways, exposing transcontinental correlations that match documented cascade patterns. In Internet networks, it reveals multiscaling behavior as the topology evolves over time.
By shifting network renormalization from spatial geometry to dynamical flow, this work provides a unified foundation for understanding how information, energy, and failures propagate through complex systems, with direct implications for infrastructure resilience and network vulnerability assessment.
\end{abstract}

\maketitle

\section*{Introduction}

Complex networks underpin diverse systems~\cite{strogatz2001exploring,albert2002statistical,newman2003structure}, ranging from brain connectivity~\cite{sporns2016networks} and protein–protein interactions~\cite{jeong2001lethality,ravasz2003hierarchical} to power grids~\cite{albert2004structural} and social networks~\cite{watts1998collective,pastor2004evolution}. These networks are vast and intricately organized. Brain networks comprise billions of neurons~\cite{sporns2016networks}, the Internet links millions of autonomous systems~\cite{pastor2004evolution}, and regulatory networks coordinate thousands of genes~\cite{lee2002transcriptional,davidson2010regulatory,alon2019introduction}. Such complexity obscures underlying organizing principles and complicates predictions of system-scale behaviour. Extracting meaningful insights therefore requires systematic coarse-graining methods~\cite{kadanoff1966scaling} that reduce complexity while preserving the essential features that govern large-scale organization. Coarse-grained representations yield simplified networks in which organizational structures become more readily identifiable.

Renormalization group (RG) theory provides a powerful framework linking microscopic interactions to macroscopic behaviour. Originally developed in statistical physics to describe critical phenomena~\cite{kadanoff1966scaling,migdal1996recursion,wilson1975renormalization,wilson1983renormalization}, RG eliminates microscopic detail while preserving the scaling laws that determine large-scale properties. Iterative coarse-graining has revealed universal principles underlying phase transitions and critical behaviour. Extending RG to complex networks, however, presents significant challenges. Scale-free networks~\cite{fox2005revisiting,barabasi2009scale}, characterized by degree distributions $P(k) \sim k^{-\gamma_d}$, also display small-world properties. Their diameters grow only sub-linearly with system size $N$ ($\ell \sim \ln N$ for $\gamma_d > 3$ and $\ell \sim \ln \ln N$ for $2 < \gamma_d < 3$)~\cite{cohen2003scale}. Such features conflict with the spatial-scale and locality assumptions of conventional RG, particularly in the regime $2 < \gamma_d < 3$, where hubs dominate connectivity.

Two main strategies have been developed to address these difficulties. Real-space RG methods, such as box-covering algorithms~\cite{song2005self,song2006origins,goh2006skeleton,kim2007fractality,kim2007fractality_pre,lepek2025beyond}, partition a network into subgraphs of diameter $\ell_B$, each treated as a renormalized supernode. This approach captures purely structural scaling. For example,
\begin{align}
N_B(\ell_B) &\sim \ell_B^{-d_B}, \quad
k'(\ell_B) \sim \ell_B^{-d_k}k,
\end{align}
where $N_B$ is the number of supernodes, $k'$ their rescaled degree, and $k$ the maximum degree within a subgraph. One finds $d_k = d_f/(\gamma_d - 1)$. Such power-law relations reveal fractality and self-similarity.

However, structural adjacency alone does not capture {\it effective interactions} in real systems. Connectivity often arises from flows of energy or information—for example, cascading failures in power grids or signal propagation in gene-regulatory networks. Explaining large-scale organization therefore requires a framework that reflects dynamical correlations rather than mere adjacency. Spectral-space RG methods, such as Laplacian RG (LRG)\cite{villegas2022laplacian,villegas2023laplacian,villegas2025multi,ghavasieh2024diversity,poggialini2025networks,nurisso2025higher,caldarelli2024laplacian,gabrielli2025network}, attempt to address this need by tracking diffusion processes. The progressive reduction of Laplacian dimensionality can be interpreted as renormalization in statistical physics. While this links RG to information propagation, in practice, coarse-graining still often depends on spatial adjacency rather than genuine dynamical correlations, leaving the central challenge unresolved\cite{villegas2023laplacian,caldarelli2024laplacian,gabrielli2025network}.

Here, we introduce a meta-graph reconstruction algorithm that derives renormalized networks directly from dimensionally reduced Laplacian matrices. By systematically translating these reduced Laplacians into explicit connectivity patterns, the algorithm produces renormalized networks that preserve both dynamical correlations and macroscopic topological structures, enabling consistent extraction of spectral, fractal, and random-walk dimensions. When applied to Internet topologies, yeast transcriptional networks, and the European power grid, our approach (i) preserves essential connectivity and topology, (ii) reveals multiscale behaviours indicative of structural transitions, and (iii) uncovers long-range dynamical correlations.

By grounding renormalization in information or energy flow rather than spatial adjacency, our meta-graph algorithm provides a systematic framework that translates dynamical correlations into structural representations. When spectral space is projected back into node space, the Laplacian RG (LRG) transformation induces correlations between structurally distant nodes, which manifest as emergent long-range links in the renormalized network. This mechanism departs fundamentally from conventional RG approaches based on spatial adjacency. As a result, the method reveals macroscopic structures—including long-range dynamical correlations—that existing frameworks fail to capture. Beyond diffusion, the approach naturally extends to a broad class of dynamical processes, enabling the development of coarse-grained models of complex systems that more faithfully reflect the interactions among their components.


\section{Analytical Approach}

To construct renormalized networks that capture dynamical correlations between nodes, our meta-graph algorithm requires a rigorous theoretical foundation that establishes the relationship between the dimensionally reduced Laplacian (which encodes these dynamical correlations) and the renormalized network. While the progressive reduction of Laplacian dimensionality during diffusion has been empirically observed, a systematic formulation within RG theory has been missing. We establish such a framework by recasting the problem in the Gaussian model, which provides the theoretical basis for systematic network renormalization.

The Gaussian model is defined by the Hamiltonian
\begin{align}
-H = \sum_{i=0}^{N-1} \left( - r \psi_{i}^{2} + K \sum_{j} \psi_{i} L_{ij} \psi_{j} \right) + \sum_i h_i \psi_i,
\end{align}
where $\psi_i$ denotes the field variable at node $i$, $L_{ij}$ is the Laplacian matrix, $K$ controls the coupling strength between nodes, and $-r \psi_i^2$ specifies local dynamics. The external field $h_i \psi_i$ probes the system’s response~\cite{kim1984random}.

A key step is the transformation from node space to spectral space. Node-based fields $\psi_i$ are expanded into spectral components $\psi_{\lambda_k}$:
\begin{align}
\psi_i = \sum_{\lambda_k} c_{i,\lambda_k} \psi_{\lambda_k},
\end{align}
with coefficients $c_{i,\lambda_k} = \langle \psi_i, \psi_{\lambda_k} \rangle$ and Laplacian eigenvalues $\lambda_k$. This diagonalization decouples spectral modes and enables systematic RG analysis.

RG calculations (see Methods) yield two fundamental scaling exponents:
\begin{align}
y_t = d_w, \quad y_h = \tfrac{1}{2}(d_f + d_w),
\end{align}
where $y_t$ and $y_h$ follow from the scaling transformations $r' = \ell^{y_t} r$ and $h' = \ell^{y_h} h$ under coarse-graining by scale factor $\ell$.

Identifying $y_t = d_w$ establishes a direct correspondence between diffusion and renormalization. Diffusion time scales as $\tau \sim \ell^{d_w} = \ell^{y_t}$, paralleling the role of $y_t$ in critical phenomena~\cite{wilson1975renormalization,wilson1983renormalization}, where it governs temperature scaling $\tau \sim (T - T_c)^{-1}$. This correspondence further implies that the dimensions governing diffusion processes map directly onto the critical exponents of critical phenomena, enabling all critical exponents to be expressed solely in terms of two network quantities—the fractal dimension $d_f$ and the random-walk dimension $d_w$. Through this scaling relation, we mathematically demonstrate that network diffusion constitutes an RG transformation in spectral space, the analogue of momentum space.

We arrive at a fundamental conclusion: the dimensionally reduced Laplacian is the Laplacian of the renormalized network. Therefore, any valid renormalized network must reflect the dynamical correlations encoded in the dimensionally reduced Laplacian. This principle yields practical guidelines: spectral information must be translated into explicit connectivity patterns that retain these correlations, providing the theoretical foundation for the construction of renormalized networks.

\subsection*{Application to Scale-Free Networks}

We apply the Gaussian model to random scale-free networks with degree distribution $P(k) \sim k^{-\gamma_d}$. Within this setting, our framework yields closed-form expressions for the three fundamental network dimensions:
\begin{align}
d_s &= \frac{2(\gamma_d - 1)}{2\gamma_d - 3}, \quad
d_f = \frac{\gamma_d - 1}{\gamma_d - 2}, \quad
d_w = \frac{2\gamma_d - 3}{\gamma_d - 2}.
\end{align}
Full derivations are provided in the Supplementary Materials.

These analytical predictions are in exact agreement with earlier results obtained from independent approaches~\cite{burda2001statistical,goh2003sandpile,havlin1987diffusion}, thereby validating our theoretical framework. This concordance demonstrates that structural and dynamical RG treatments, though originating from distinct perspectives, ultimately converge to the same scaling laws for scale-free networks.


\begin{figure*}[!t]
\centering
\includegraphics[width=1.0\linewidth]{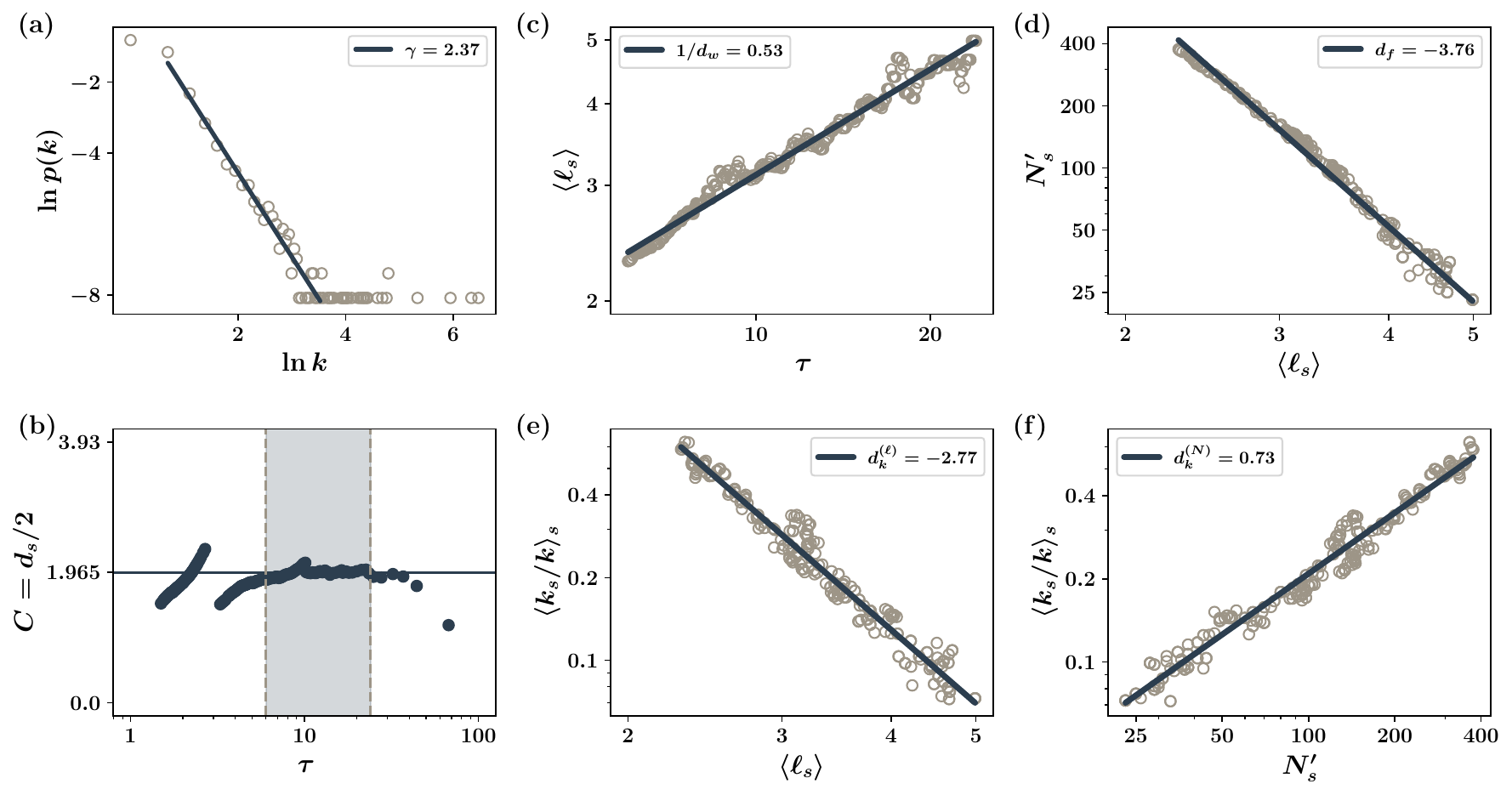}
\caption{
\textbf{Scaling analysis of Internet topology.} 
Application of our meta-graph algorithm to Internet AS-level topology (1998) reveals complete theoretical consistency across all fundamental network dimensions.
(a) Network exhibits power-law degree distribution $P(k) \sim k^{-2.37}$ for $2 \leq k \leq 36$.
(b) Spectral dimension from specific heat plateau: $C = d_s/2 = 1.965$ for $\tau \in [6, 24]$ yields $d_s = 3.93$.
(c) Random-walk dimension from supernode size scaling: slope $1/d_w = 0.53$ gives $d_w = 1.89$.
(d) Fractal dimension from supernode number scaling: $N_s'$ vs $\langle \ell_s \rangle$ yields $d_f = 3.77$.
(e) Degree scaling validation: $\langle k_J' / k \rangle_J$ vs $\langle \ell_s \rangle$ gives $d_k^{(\ell)} = 2.79$.
(f) Theoretical consistency: measured $d_k/d_f = 0.74$ matches prediction $1/(\gamma_d - 1) = 0.73$.
All independently measured dimensions satisfy the Alexander--Orbach relation ($d_f/d_w = 1.99 \approx d_s/2 = 1.98$), confirming our comprehensive approach captures both structural and dynamical scaling within a single consistent framework.
Additional supernode distributions in Supplementary Figs.~\ref{fig:internet_degree}--\ref{fig:internet_size}.
}
\label{fig:internet}
\end{figure*}

\section{Numerical Results}

\subsection{Constructing the Binary Renormalized Network}

Building on our theoretical framework—which established that renormalized networks must reflect the correlations encoded in dimensionally reduced Laplacian—we develop a meta-graph reconstruction algorithm and apply it to real-world networks across diverse domains. This approach yields renormalized networks that faithfully capture the dynamical correlations inherent in the original systems.

A central challenge lies in translating renormalized spectral information back into an interpretable network topology. Unlike conventional renormalization methods in Euclidean settings, such as momentum- or real-space RG, this reverse transformation is indispensable for complex networks. Because node space is highly heterogeneous, the mapping depends critically on both the degree of coarse-graining performed in spectral space and the choice of projection matrix.

To address this challenge, we introduce a meta-graph reconstruction algorithm that embeds dynamical correlations directly into structural connectivity through spectral clustering. Using a projection matrix, nodes with similar responses to perturbations are merged into supernodes, while their interconnections are defined exclusively by diffusion dynamics. In this way, coarse-graining in eigenspace is systematically translated into a renormalized topology, with information-flow pathways explicitly encoded as structural links.

Crucially, our framework operates across multiple time scales, generating dynamical–structural representations at each level. As time scales vary, both diffusion and the resulting structures evolve. For coherence, the diffusion process—governed by the random-walk dimension $d_w$—must remain consistent with the renormalized network—governed by the fractal dimension $d_f$. This consistency is enforced by the Alexander–Orbach (AO) relation, $d_s = 2d_f/d_w$~\cite{alexander1982density,orbach1986dynamics}, ensuring that renormalized networks preserve the balance between dynamics and structure throughout the multiscale framework.

Our approach thus enables the extraction of network dimensions that remain valid across scales and reveals large-scale organizations inaccessible to conventional RG methods based solely on spatial adjacency.

We demonstrate the broad applicability of this framework through systematic analyses of real-world systems, including Internet topologies, biological regulatory networks, and critical infrastructure. Detailed algorithmic procedures are provided in the Methods section (see Fig.~\ref{fig:flowchart_ds_procedure}).

\subsection{Scaling Relations in Real Networks}

To evaluate our framework, we apply the meta-graph reconstruction algorithm to diverse real-world networks spanning technological, biological, and infrastructure domains. This analysis demonstrates both the robustness and the broad applicability of our approach across fundamentally different architectures.

We first examine the Internet topology at the autonomous system (AS) level, one of the most complex and extensively studied technological infrastructures. Analysis of the CAIDA dataset (199801)\cite{caida_as}, which follows a power-law degree distribution $P(k) \sim k^{-2.37}$, illustrates the core capability of our framework. From this network, we obtain three key dimensions: $d_s = 3.93$, $d_w = 1.89$, and $d_f = 3.77$. These values remain stable across diffusion times $\tau \in [6, 24]$ [Fig.\ref{fig:internet}]. All scaling relations are satisfied with high precision, as shown by the agreement between the measured ratio $d_k/d_f = 0.74$ and the theoretical prediction $1/(\gamma_d - 1) = 0.73$. This consistency confirms that our renormalization procedure preserves both connectivity and scaling structure throughout coarse-graining, further supported by the preservation of the degree distribution [Fig.~\ref{fig:internet_degree}].

The versatility of the framework is further highlighted in networks with very different characteristics. The yeast transcriptional regulatory network~\cite{teixeira2006yeastract}, representing biological regulation with degree exponent $\gamma_d = 1.77$, yields $d_s = 6.12$, $d_f = 2.82$, and $d_w = 0.92$. The European power grid~\cite{pagnier2019inertia}, an engineered infrastructure with an exponential rather than power-law degree distribution, gives $d_s = 2.00$, $d_f = 1.95$, and $d_w = 1.96$. In both cases, all scaling relations hold with high accuracy [Fig.~\ref{fig:yeast}, \ref{fig:powergrid}], confirming that our method extracts consistent scaling behavior even in structurally diverse systems.

Together, these results establish a methodological advance: our framework simultaneously determines all network dimensions governing collective behavior while ensuring complete theoretical consistency. By integrating structural properties ($d_f$) with dynamical properties ($d_s$, $d_w$) in a single analysis, it provides a comprehensive characterization of network scaling that spans from architecture to transport processes. This represents a significant improvement over existing approaches, which typically estimate individual dimensions in isolation.

\begin{figure*}[!t]
\centering
\includegraphics[width=1.0\linewidth]{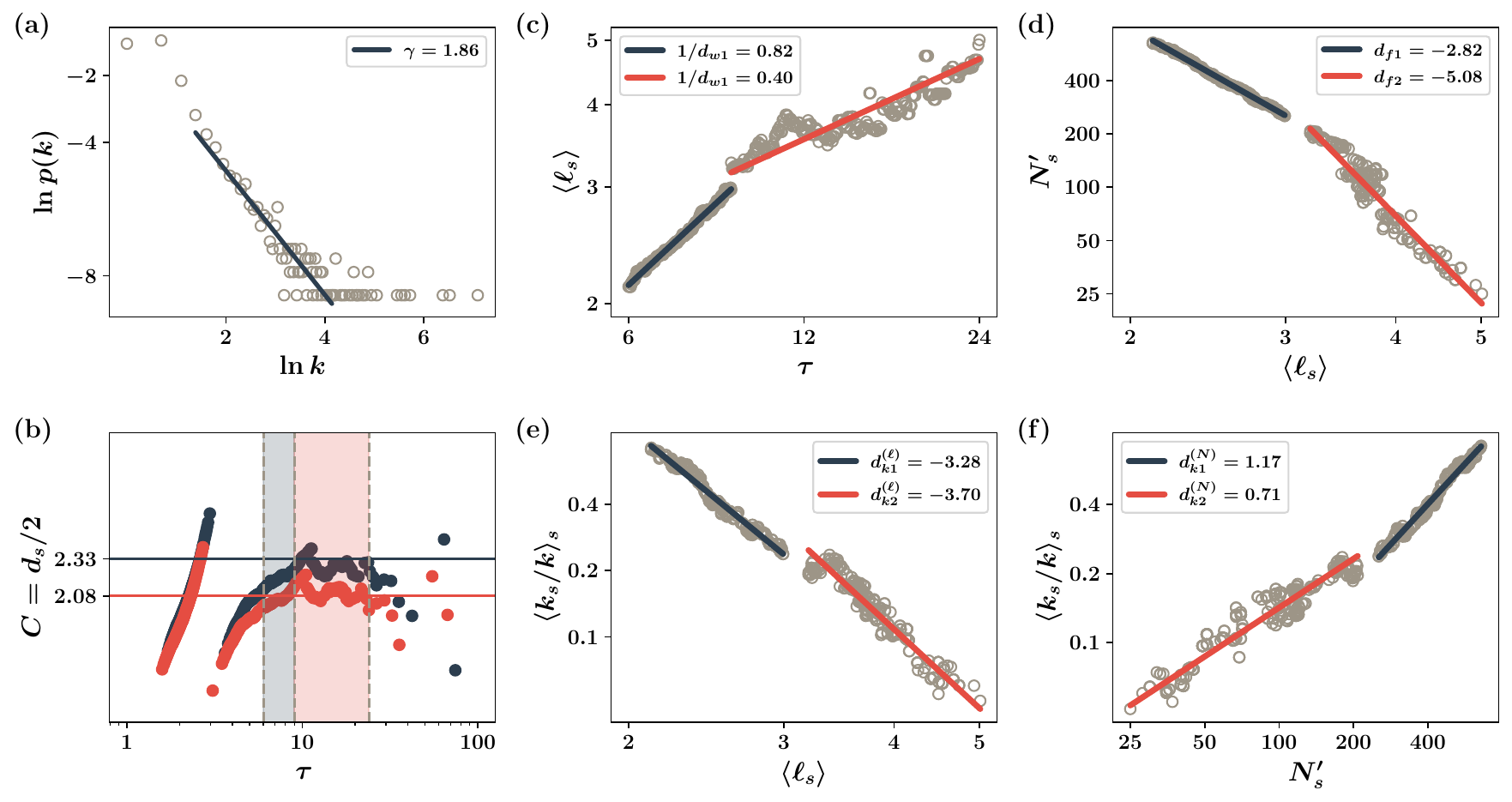}
\caption{
\textbf{Multi-scaling behavior in evolving networks.} 
Application of our meta-graph algorithm to Internet AS-level topology (1999) demonstrates crossover between distinct scaling regimes, revealing coexisting network substructures.
(a) Degree distribution with $\gamma_d \approx 1.86$ for $k \in [4, 66]$ shows heavier tail compared to 1998 data.
(b) Spectral dimension crossover: $C = d_s/2 = 2.33$ (short $\tau \in [6, 9]$) vs $2.08$ (long $\tau \in [9, 24]$).
(c) Random-walk dimension regime change: $1/d_w = 0.82$ (short $\tau$) vs $0.40$ (long $\tau$) indicates different transport mechanisms.
(d) Fractal dimension transition: $d_f \approx 2.82$ (short $\tau$) vs $5.08$ (long $\tau$) reflects structural reorganization.
(e) Degree scaling consistency within regimes: $d_k^{(\ell)} \approx 3.28$ (short $\tau$) and $3.70$ (long $\tau$).
(f) Theoretical validation: $d_k^{(\ell)}/d_f = 1.17 \approx d_k^{(N)} = 1.17$ matches $1/(\gamma_d - 1) = 1.17$.
The Alexander--Orbach relation remains valid in both regimes despite different exponent values, confirming that multiple coexisting scaling behaviors challenge universal network scaling assumptions.
}
\label{fig:internet_1999}
\end{figure*}

\subsection{Multi-Scaling Behavior in Network Evolution}

Real-world networks evolve over time, potentially reshaping their scaling properties and structural organization. To assess these effects, we apply our framework to Internet topologies at multiple time points and uncover pronounced multi-scaling behavior that challenges the assumption of a single universal scaling regime.

Analysis of AS-level Internet datasets (199907 and 200101) shows that these networks do not obey a uniform set of scaling rules. Instead, they exhibit coexisting substructures: hub-dominated cores and scale-free peripheries, each governed by distinct scaling behaviors. This demonstrates that our framework can not only extract network dimensions but also diagnose structural heterogeneity and regime transitions.

The multi-scaling emerges as a crossover in network dimensions depending on the diffusion time scale $\tau$ [Figs.~\ref{fig:internet_1999}, \ref{fig:internet_2001}(c)–(f)]. Two distinct regimes appear:

\textbf{Short-time regime} (small $\tau$): The degree–fractal dimension relation $d_k/d_f$ remains consistent with the theoretical prediction $1/(\gamma_d - 1)$, reflecting the scale-free periphery that dominates global connectivity.

\textbf{Long-time regime} (large $\tau$): Although the fundamental relations among $d_s$, $d_f$, and $d_w$ persist, degree scaling breaks down. The high-degree core produces a much broader tail than scale-free theory predicts, approaching a nearly uniform form that alters the scaling rules.

This crossover arises from the heterogeneous degree distribution of real networks. Short times probe the scale-free periphery, whereas long times reveal the influence of dense, high-degree cores where scale-free regularities fail. The persistence of multi-scaling across different temporal snapshots indicates that it reflects intrinsic structural features rather than transient fluctuations.

These findings suggest that many real-world networks possess a hybrid organization, combining scale-free peripheries with nearly uniform cores. Such multi-scale structure fundamentally modifies transport dynamics and system responses, producing effects that single-scaling approaches overlook. Our framework thus provides a powerful tool for uncovering hidden structural complexity and identifying regime transitions in evolving networks.

\subsection{Non-recursive Nature and Long-Range Dynamical Correlations}

Our meta-graph algorithm reveals a property absent from conventional RG approaches: its inherently non-recursive nature. In standard RG, iterative coarse-graining allows renormalized states to be reached through multiple pathways. By contrast, our method exhibits strong path dependence: different choices of the diffusion time $\tau$ generate distinct renormalized structures that cannot be transformed into one another by further iteration.

\begin{figure*}[!t]
\centering
\includegraphics[width=1.0\linewidth]{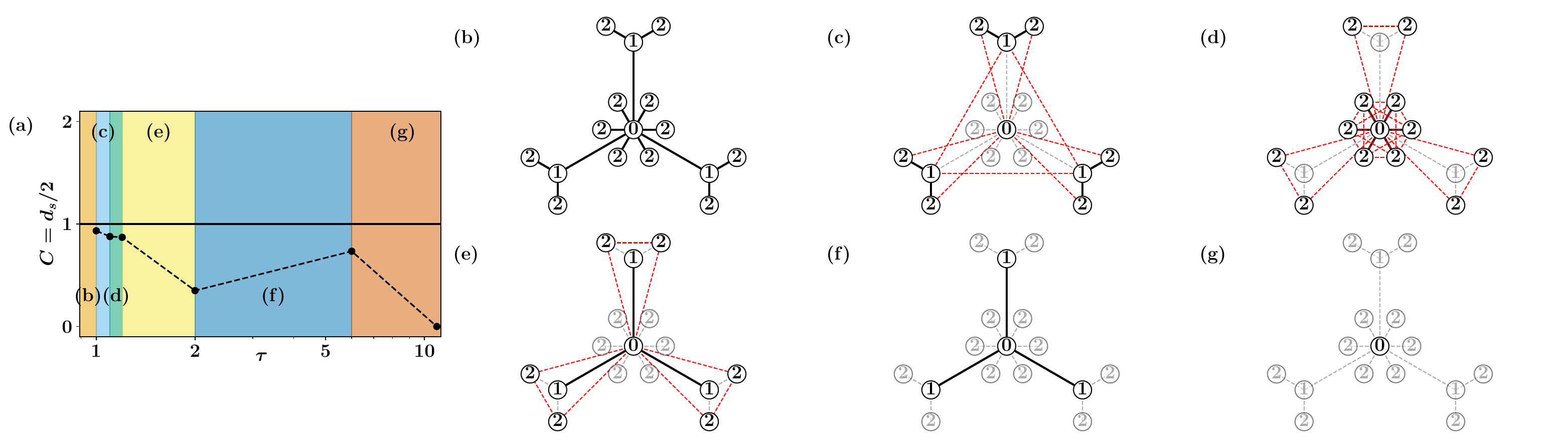}
\caption{
\textbf{Non-recursive nature of the meta-graph algorithm in deterministic networks.} 
Analysis of a three-generation deterministic scale-free tree demonstrates path dependence and irreversibility inherent in spectral space transformations.
(a) Diffusion time domains mapped to accessible renormalized topologies shown in panels (c)--(g).
(b) Original network with hierarchical tree structure across three generations.
(c)--(g) Renormalized meta-graphs at different diffusion times $\tau$ show emergence of new connections (thick red edges) and structural reorganization. Critically, specific configurations (e.g., panel f) become inaccessible from intermediate renormalized states (e.g., panel d), even though they remain directly reachable from the original network.
This path dependence demonstrates the fundamental non-recursive and irreversible nature of the meta-graph algorithm, distinguishing it from conventional geometric coarse-graining methods.
}
\label{fig:SF_tree}
\end{figure*}

This behavior is evident in deterministic scale-free networks~\cite{jung2002geometric}. At intermediate $\tau$, RG produces topological dead-ends—configurations that block access to certain renormalized states, even though those remain directly accessible from the original network (Figs.~\ref{fig:SF_tree}, \ref{fig:SF_loop}, Supplementary Materials). The origin lies in eigenspace coarse-graining, a global operation that reorganizes large-scale structure beyond local topology.

The global deformation of eigenspace provides unique analytical leverage. It uncovers organizational patterns invisible to local connectivity, systematically exposing latent dynamical correlations—regions structurally distant yet dynamically coherent under perturbations.

\begin{figure*}[!t]
\centering
\includegraphics[width=1.0\linewidth]{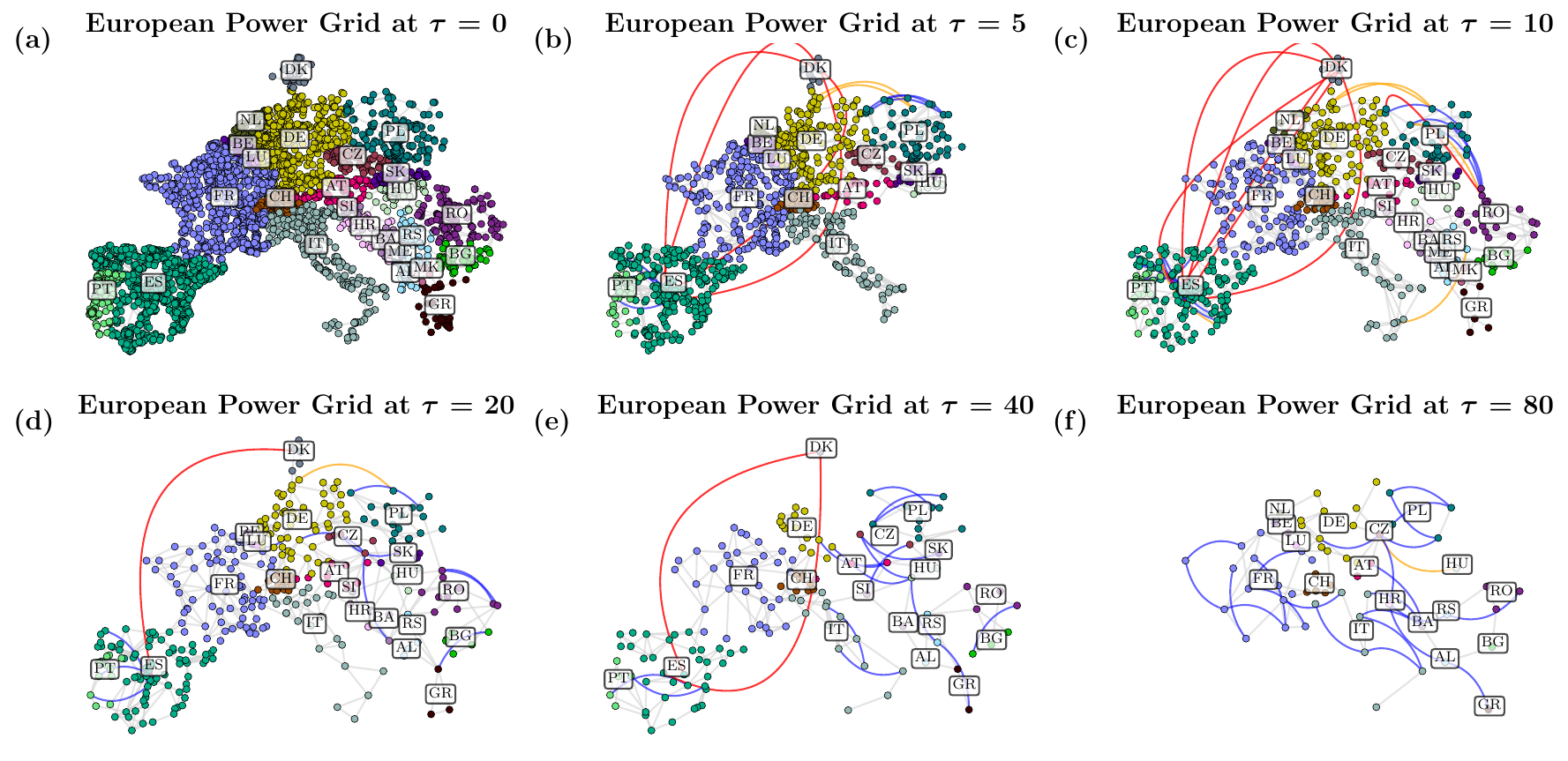}
\caption{
\textbf{Hidden dynamical correlations.} 
Meta-graph analysis of the European power grid reveals latent long-range correlations that match documented transcontinental cascade patterns.
(a) Original geographical network topology showing only local physical connections.
(b)--(f) Renormalized meta-graphs at increasing diffusion times $\tau$ consistently reveal new long-range correlations (thick red edges), most prominently between Denmark and Spain.
These dynamically coherent connections, absent from the physical topology, reflect shared spectral signatures that indicate synchronized responses to network perturbations. The Denmark-Spain correlation matches empirically observed cascade pathways where faults originating in Greece propagated through Northern Europe to destabilize the Spanish grid, validating our framework's predictive capability for infrastructure vulnerability assessment.
}
\label{fig:meta_euro_powergrid}
\end{figure*}

We demonstrate this with the European power grid, where non-recursive transformation uncovers hidden vulnerability patterns. The meta-graph reconstruction identifies long-range correlations between geographically distant regions, most notably Denmark and Spain [Figs.~\ref{fig:meta_euro_powergrid}(b)--(f)]. Despite lacking direct physical connections, these regions exhibit nearly identical spectral signatures, signaling coherent dynamical responses.

Strikingly, this prediction aligns with empirical observations: Pagnier et al.~\cite{pagnier2019inertia} reported cascade failures that began in Greece and destabilized Spain via Northern Europe, precisely reflecting our Denmark–Spain correlation. Our framework reproduces such transcontinental pathways directly from structural and dynamical properties, without relying on specific failure data.

Thus, the non-recursive nature of the meta-graph algorithm establishes it as a diagnostic tool for critical infrastructure. By identifying dynamically coherent regions in advance, it anticipates cascade pathways, informs preventive strategies, and guides stability control. Beyond topology-based analyses, this capability reveals hidden dynamical structures that govern large-scale network behavior.


\section{Conclusion and Discussion}

Renormalization group (RG) theory\cite{kadanoff1966scaling,migdal1996recursion,wilson1975renormalization,wilson1983renormalization} has long provided a unifying framework in physics, offering deep insights into critical phenomena and quantum field theory. Extending this framework to networks, however, has remained a persistent challenge. The recently proposed Laplacian RG\cite{villegas2023laplacian,caldarelli2024laplacian,gabrielli2025network} suggested a promising path by reconciling small-world properties with scale-free fractality, yet its foundation has largely remained phenomenological. This situation recalls the early development of conventional RG, where Kadanoff’s scaling intuition preceded Wilson’s formalism. Networks now face a similar turning point: recognizing renormalization in Laplacian eigenspace opens a systematic program, which we advance through a rigorous and scalable spectral-space renormalization group (SSRG) framework, supported by both theoretical analysis and empirical validation.

Unlike traditional systems, networks cannot be disentangled from the dynamics flowing along their links. Our framework captures this inseparability by deriving renormalized structures directly from dynamical correlations, ensuring coherence between topology and dynamics. The key observation is that dimensional reduction of the Laplacian during diffusion parallels the reduction of degrees of freedom near critical points in statistical mechanics. This correspondence shows that all network scaling properties can be expressed in terms of the fractal dimension $d_f$ and the random-walk dimension $d_w$. It provides a theoretical basis for interpreting diffusion phenomena within the RG framework and for deriving renormalized networks directly from dimensionally reduced Laplacian matrices.

We operationalize this principle through a meta-graph reconstruction algorithm that extracts network dimensions $d_f$, $d_s$, and $d_w$ from real networks with internal consistency. Applications to the Internet, power grids, and biological systems demonstrate robustness and universality, yielding renormalized networks that encode genuine dynamical correlations. This approach uncovers macroscopic organizational structures that remain hidden under purely structural treatments, including multiscaling behaviors and long-range meta-connections. By embedding dynamical correlations directly into topology, abstract processes become tangible structural signatures—offering practical insights for identifying failure pathways in infrastructure, mapping neural activity, tracing contagion routes, and diagnosing systemic risks in finance.

Looking forward, extending SSRG to time-varying~\cite{holme2012temporal}, multilayer~\cite{kivela2014multilayer,boccaletti2014structure}, and weighted networks~\cite{newman2004analysis} will more fully capture real-world complexity. Incorporating nonlinear, oscillatory, and multiscale dynamics promises to reveal how structure and function co-evolve~\cite{boccaletti2006complex,arenas2008synchronization,dorfler2014synchronization}. SSRG thus provides a new foundation for resilience design, from mitigating cascading failures~\cite{watts2002simple,buldyrev2010catastrophic} in power grids to understanding neural oscillations~\cite{buzsaki2006rhythms}, ecosystem stability~\cite{may1972will}, and economic dynamics~\cite{schweitzer2009economic}. The central message is clear: in networks, structure cannot be separated from dynamics but must be derived from it. Embedding dynamical correlations directly into connectivity exposes hidden large-scale organization and provides a unified language for explaining how order emerges in nature’s most intricate systems.



    

\begin{acknowledgments}
B.K. was supported by the National Research Foundation of Korea by Grant No. RS-2023-00279802 and the KENTECH Research Grant No. KRG-2021-01-007.
During manuscript preparation, we utilized Claude and ChatGPT to enhance language clarity and readability in the text. All AI-assisted content was carefully reviewed and validated by the authors, who maintain full responsibility for the scientific content, accuracy, and integrity of this work.
\end{acknowledgments}






\bibliographystyle{apsrev4-2}
\bibliography{CSF_Ref}

\clearpage
\newpage

\appendix

\onecolumngrid


\makeatletter
\renewcommand{\thesection}{S\arabic{section}}
\renewcommand{\theequation}{S\arabic{equation}}
\renewcommand{\thefigure}{S\arabic{figure}}
\renewcommand{\thetable}{S\arabic{table}}
\renewcommand{\bibnumfmt}[1]{[S#1]}

\setcounter{section}{0}
\setcounter{equation}{0}
\setcounter{figure}{0}
\setcounter{table}{0}
\setcounter{page}{1}
\makeatother

\begin{center}
\textbf{\large Supplementary Material for\\ ``From Spatial to Spectral: Network Renormalization via Dynamical Correlations"}
\end{center}

\section{Methods}


\subsection{Detailed Renormalization Group Procedure}

This section outlines the mathematical foundation of our spectral space RG (SSRG) framework. Starting from the network Laplacian, we construct the Gaussian model and apply renormalization group (RG) techniques to derive scaling dimensions and establish relationships governing network behavior across scales.

\subsubsection{Spectral Transformation}

The essential step is transforming the Gaussian model from node space to eigenspace. In node space, the Laplacian encodes structural connectivity, whereas in eigenspace it naturally represents diffusion dynamics.

We expand node-based fields $\{\psi_i\}$ in terms of eigenmodes $\{\psi_{\lambda_k}\}$ of $L_{ij}$:
\begin{align}
\psi_i = \sum_{\lambda_k} c_{i,\lambda_k} \psi_{\lambda_k},
\end{align}
where $c_{i,\lambda_k}$ are Laplacian eigenvectors. On this basis, the Hamiltonian (with $h_i=0$) diagonalizes:
\begin{align}
-H_\lambda = - r \sum_{\lambda_k} \psi_{\lambda_k}^2 + K \sum_{\lambda_k} \lambda_k \psi_{\lambda_k}^2.
\end{align}
This decoupling of spectral modes renders the RG analysis tractable.

\subsubsection{Four-Step RG Transformation}

Our SSRG procedure follows the standard four steps of statistical physics, adapted to networks.

\paragraph{Step 1: Coarse-graining}
The spectrum is partitioned at cutoff $\lambda_*$ into low-frequency modes $\psi^<_{\lambda}$ ($\lambda<\lambda_*$) and high-frequency modes $\psi^>_{\lambda}$ ($\lambda>\lambda_*$). Low modes capture slow, collective processes, while high modes represent fast, local fluctuations. The low-frequency Hamiltonian reads:
\begin{align}
-H_\lambda^< = -r \sum_{\lambda_k < \lambda_*} \psi_{\lambda_k}^2 
+ K \sum_{\lambda_k < \lambda_*} \lambda_k \psi_{\lambda_k}^2.
\end{align}

\paragraph{Step 2: Integrating out high-frequency modes}
High-frequency contributions are integrated out, yielding the effective partition function
\begin{align}
Z_\lambda = \int \prod_{\lambda_k < \lambda_*} d\psi_{\lambda_k} 
\exp \!\left[-r \sum \psi_{\lambda_k}^2 + K \sum \lambda_k \psi_{\lambda_k}^2 \right].
\end{align}
This step removes microscopic fluctuations while retaining large-scale behavior.

\paragraph{Step 3: Rescaling}
We rescale eigenvalues and fields as $\lambda \to \lambda'/b$ and $\psi_\lambda \to \zeta \psi'_{\lambda'}$, with $b=\lambda_{\max}/\lambda_*$ and $\zeta^2=b^{d_s/2+1}$, where $d_s$ is the spectral dimension. Scale invariance requires
\begin{align}
r' = rb = r\ell^{y_t}, \quad K'=K.
\end{align}
The thermal exponent follows as $y_t = \ln b/\ln \ell = d_w$, linking spectral scaling to random-walk dynamics.

\paragraph{Step 4: External field scaling}
An external perturbation couples to the zero mode: $H_\lambda \to H_\lambda + h \psi_{\lambda=0}$. Under RG this field rescales as
\begin{align}
h \to h' = h b^{1/2+d_s/4} = h \ell^{y_t/2+d_f/2} \equiv h \ell^{y_h},
\end{align}
using the Alexander–Orbach relation $d_s/2 = d_f/d_w$. This yields the magnetic exponent
\[
y_h = \frac{d_f+y_t}{2},
\]
governing the network’s response to external perturbations.

\subsubsection{Connection to Critical Exponents}

To complete our theoretical framework and highlight its broader significance, we establish how the fundamental network dimensions $(d_f, d_s, d_w)$ connect to the well-established critical exponents $(\alpha, \beta, \gamma)$ in statistical mechanics through the thermal and magnetic exponents, $y_t = d_w$ and $y_h = (d_f + d_w)/2$. This correspondence reveals a deep structural link between network science and critical phenomena.

The mapping between network dimensions and classical critical exponents is given by
\begin{align}
\alpha = 2 - \frac{d_f}{y_t}, \quad
\beta = \frac{d_f - y_h}{y_t}, \quad
\gamma = \frac{2y_h - d_f}{y_t}.
\end{align}

These relations build a direct bridge between diffusion processes in complex networks and phase transitions in condensed matter systems. The correspondence is not merely formal: it emerges through the diffusion time, $\tau_* \equiv \lambda_{\rm max}/\lambda_*$, which sets the characteristic scale for a random walker to traverse a region of size $\ell_* \sim \tau_*^{1/d_w}$. In this analogy, $\tau$ plays the role of an inverse temperature: just as temperature controls the number of thermally accessible degrees of freedom, $\tau$ determines the number of spectral modes contributing to the network’s effective dynamics through the eigenvalue cutoff $\lambda_*$.

This perspective suggests that scale-invariant networks can be understood as critical systems, exhibiting critical-like behavior in their stability, susceptibility to perturbations, and dynamical evolution. Accordingly, our framework not only provides a systematic method to extract network dimensions but also situates complex networks within the universality class of critical phenomena, offering a unified language for interpreting their large-scale organization and dynamic behavior.

\clearpage
\newpage


\subsection{Detailed Derivation for Random Scale-Free Networks}

We now apply our framework—based on the Gaussian model and SSRG analysis—to random scale-free networks, which provide analytically tractable test cases. Explicit expressions for the network dimensions ($d_f$, $d_s$, $d_w$) are derived as functions of the degree exponent $\gamma_d$, and shown to coincide with results obtained from independent approaches, thereby validating our theory.

\subsubsection{Spectral Dimension}

From box-covering RG methods~\cite{song2005self,song2006origins,cohen2010complex}, the maximum degree scales under coarse-graining as
\begin{align}
\frac{k'_{\max}}{k_{\max}} \sim \left( \frac{N'}{N} \right)^{1/(\gamma_d - 1)},
\end{align}
where $k_{\max}$ and $k'_{\max}$ denote maximum degrees before and after RG. Since the number of supernodes scales as $N'/N \sim \ell^{-d_f}$ with length scale $\ell$, this implies $k' \sim \ell^{-d_k}k$ with $d_k = d_f/(\gamma_d - 1)$~\cite{song2005self}.

In parallel, SSRG predicts eigenvalue scaling $\lambda' = \lambda/b$ with $b \sim \ell^{y_t}$. Within the Gaussian model, this corresponds to $k' \sim \ell^{-2d_f+y_t}k$ (see detailed derivation below). Consistency of the two approaches requires
\begin{align}
-2d_f + y_t = -d_k.
\end{align}
Substituting $d_k = d_f/(\gamma_d - 1)$ and using the Alexander–Orbach relation $d_s = 2d_f/y_t$ yields
\begin{align}
d_s = \frac{2(\gamma_d - 1)}{2\gamma_d - 3}.
\end{align}
This expression holds for $2 < \gamma_d < 3$, reproducing $d_s=2$ at $\gamma_d=2$ and $d_s=4/3$ at $\gamma_d=3$, consistent with earlier studies~\cite{burda2001statistical}. The limiting case $d_s=4/3$ echoes the Alexander–Orbach conjecture for anomalous diffusion in disordered media.

\subsubsection{Fractal Dimension}

To obtain $d_f$, we use the Einstein relation for diffusion on fractals, $d_w = d_f + \tilde{\zeta}$~\cite{havlin1984relation}.  
For random scale-free networks, the tree-like branching structure implies $\tilde{\zeta}=1$, giving $d_w = d_f+1$. Combining with $d_w=2d_f/d_s$ and the above result for $d_s$, we find
\begin{align}
d_f = \frac{\gamma_d - 1}{\gamma_d - 2}.
\end{align}
This matches independent results, including those from sandpile dynamics~\cite{goh2003sandpile}.

\subsubsection{Complete Scaling Relations}

Having determined $d_s$ and $d_f$, the full set of exponents follows:
\begin{align}
y_t = d_w = \frac{2d_f}{d_s} = \frac{2\gamma_d - 3}{\gamma_d - 2}, 
\qquad 
y_h = \frac{d_f + y_t}{2} = \frac{3\gamma_d - 4}{2(\gamma_d - 2)}.
\end{align}
Thus, all scaling exponents can be expressed directly in terms of $\gamma_d$, providing a complete analytical characterization of random scale-free networks.

\clearpage
\newpage


\subsection{Deriving the Scaling Relation for $d_k$ in Random Scale-Free Networks}
\label{sec:deriving_scaling_relation}

We now derive the scaling relation,
\[
2d_f - y_t = d_k,
\]
by requiring that two different renormalization schemes—the box-covering method and the spectral-space renormalization group (SSRG)—yield consistent results when applied to the same network. This condition provides a quantitative link between geometry, dynamics, and topology.

\begin{enumerate}
    \item In random scale-free networks, the critical branching property ensures that the degree of the dominant hub within a box, $k_{\rm hub}$, is proportional to the number of external links from that box, $k_{\rm Box}$. Owing to the tree-like structure of such networks, cycles are rare, and the external connectivity of a box scales with the branching of its hub. Thus, $k_{\rm Box}$ is effectively governed by $k_{\rm hub}$, reflecting the hierarchical organization that underpins our analysis.
    
    \item The probability that two boxes $I$ and $J$ are connected can be expressed from the microscopic viewpoint as
    \[
    P(I,J) \propto \frac{1}{N \langle k \rangle} \left( \sum_{i \in I} k_i \right) \left( \sum_{j \in J} k_j \right).
    \]
    Because random scale-free networks are statistically homogeneous, local averages satisfy $\langle k_i \rangle_I = \langle k_j \rangle_J = \langle k \rangle$, giving
    \[
    P(I,J) \propto \frac{\langle k \rangle}{N}\, |I|\, |J|,
    \]
    where $|I|$ and $|J|$ are the number of nodes in boxes $I$ and $J$, respectively.
    
    \item Within the SSRG framework, coarse-graining rescales both the effective degree and box size:
    \[
    P(I,J) \;\;\rightarrow\;\;
    \frac{b \langle k \rangle}{\ell^{-d_f} N}\,
    (\ell^{-d_f}|I|)(\ell^{-d_f}|J|)
    = \frac{\ell^{-2d_f} b}{\ell^{-d_f}} P(I,J),
    \]
    where $b = \ell^{y_t}$ captures the rescaling of inter-box connectivity and $\ell^{-d_f}$ accounts for the reduced number of nodes per box.
    
    \item Alternatively, viewing each box as a supernode with degree $k_{\rm Box}$, the connection probability becomes
    \[
    P(I,J) \propto \frac{k_I k_J}{N_{\rm Box} \langle k_{\rm Box} \rangle}
    \;\;\rightarrow\;\;
    \frac{\ell^{-d_k} k_I \cdot \ell^{-d_k} k_J}{\ell^{-d_f} N_{\rm Box}\,\ell^{-d_k}\langle k_{\rm Box} \rangle}
    = \frac{\ell^{-d_k}}{\ell^{-d_f}} P(I,J).
    \]
    Here, the scaling factor $\ell^{-d_k}$ arises because $k_{\rm Box}$ is proportional to $k_{\rm hub}$ under the critical branching condition.
    
    \item Demanding consistency between the two renormalization perspectives yields
    \[
    \frac{\ell^{-2d_f} b}{\ell^{-d_f}} P(I,J)
    = \frac{\ell^{-d_k}}{\ell^{-d_f}} P(I,J),
    \]
    which, with $b = \ell^{y_t}$, reduces to
    \[
    \ell^{-2d_f+y_t} = \ell^{-d_k},
    \]
    giving
    \[
    2d_f - y_t = d_k.
    \]
\end{enumerate}

This scaling law reveals that the fractal dimension $d_f$, the dynamical exponent $y_t=d_w$, and the degree exponent $d_k$ are not independent quantities but interdependent outcomes of a common renormalization symmetry. It thereby unifies structural complexity, dynamical efficiency, and topological organization of scale-free networks within a single framework.

\clearpage
\newpage


\subsection{Construction of the Renormalized Network (Meta-Graph)}
\label{sec:Construction_of_Renormalized_Network}

We describe the meta-graph reconstruction algorithm, which bridges spectral-space analysis with node-space topology while retaining both structural and dynamical information. This transformation enables the direct application of our framework to empirical networks, linking abstract spectral analysis with practical network characterization.

The key principle is that Laplacian eigenvectors encode not only structural connectivity but also dynamical coherence among nodes. By analyzing these spectral signatures, we group nodes with similar dynamical responses into supernodes, while their interconnections define the links of the meta-graph. The resulting network preserves scaling properties while reducing complexity, offering a faithful multiscale representation of the original system.

\begin{enumerate}
\item[1.] \textbf{Spectral Mode Selection}

\textit{Objective:} Select relevant spectral modes and form a projection matrix encoding effective dynamical couplings.

We start with the Laplacian $L_{ij}$, where $L_{ij}=-1$ if nodes $i$ and $j$ are connected, $L_{ij}=0$ otherwise, and $L_{ii}=-\sum_{j \neq i} L_{ij}$. The coarse-graining scale is set by the diffusion time $\tau$. Eigenvalues $\lambda_k$ are retained if $\lambda_k < \lambda_{*}/\tau$, with $\lambda_{*}$ the largest eigenvalue. The projection matrix is then
\begin{align}
\mathcal{P} = \sum_{\lambda_k < \lambda_{*}/\tau} | \lambda_k \rangle \lambda_k \langle \lambda_k |,
\end{align}
whose node-basis elements $a_{ij}$ quantify dynamical coupling between $i$ and $j$.

\item[2.] \textbf{Node Classification}

\textit{Objective:} Distinguish supernodes from absorbed nodes via spectral coherence.

A node $i$ is absorbed if its self-coupling is weaker than its strongest external coupling:
\[
\mathcal{P}_{ii} < \max_j (\mathcal{P}_{ij}).
\]
Absorbed nodes form category $\text{C}_n$, while surviving nodes form category $\text{C}_s$. This spectral criterion groups nodes by dynamical coherence rather than geometric proximity.

\item[3.] \textbf{Supernode Assignment and Connectivity Transfer}

\textit{Objective:} Assign absorbed nodes to supernodes and update interconnections.

Each absorbed node is assigned to the supernode with maximum cosine similarity in spectral space:
\[
\cos \theta_{Ij} = \frac{\mathcal{P}(I)\cdot \mathcal{P}(j)}{\|\mathcal{P}(I)\|\|\mathcal{P}(j)\|}.
\]
Links are then transferred: if $j \to e$, absorbed into $J$ and $E$, the connection becomes $J \to E$, with weights aggregated as
\[
a_{JE} = \sum_{j \in J, \, e \in E} a_{je}.
\]

\item[4.] \textbf{Network Binarization}

\textit{Objective:} Convert the weighted meta-graph into a standardized Laplacian form.

Dominant off-diagonal couplings are retained, binarized ($a_{IJ}=-1$), and diagonals adjusted to enforce flow conservation. This yields a simplified Laplacian preserving the strongest dynamical pathways.

\item[5.] \textbf{Finalization}

Disconnected components are merged by reassigning nodes via cosine similarity. The final meta-graph captures both structural and dynamical organization in a compact form.
\end{enumerate}

\textbf{Algorithm Summary:}  
The meta-graph algorithm systematically reconstructs a renormalized topology from spectral information. Unlike geometric or static adjacency methods, it derives connections from effective dynamical correlations. As a result, the meta-graph reflects not only local connectivity but also coherent dynamical response, providing an interpretable framework for multiscale network analysis.

\clearpage
\newpage

\subsection{Construction of Renormalized Network---Flow Chart}
\label{sec:Flow_Chart}

\begin{figure*}[!htbp]
\centering
\begin{adjustbox}{width=\textwidth}
\begin{tikzpicture}[
  every node/.style={font=\Large},
  font=\footnotesize,
  node distance = 15mm and 20mm,            
  process/.style = {rectangle, draw, align=center, minimum width=28mm, minimum height=6mm},
  decision/.style = {diamond,   draw, aspect=2, align=center, inner sep=0.7pt, minimum width=26mm},
  terminator/.style = {ellipse,  draw, align=center, minimum width=24mm, minimum height=6mm},
  arrow/.style = {-{Latex[length=2.3mm]}, line width=0.45pt, shorten >=1pt}
]

\node[terminator] (start) {Start};

\node[process, below=of start] (init) {Initial setting\\($d_s/2$, $[\tau_1,\tau_2]$, $\lambda_{\max}$)};

\node[process, below=of init] (selectEig) {For each $\tau\in[\tau_1,\tau_2]$\\select $\{\lambda_k\le\lambda_{\max}/\tau\}$};

\node[process, below=of selectEig] (avgEig) {Compute $\langle\lambda_k\rangle(\tau)$ for each $\tau\in[\tau_1,\tau_2]$};

\node[process, below=of avgEig] (avgInterval) {Compute $C_*=\overline{\tau\langle\lambda_k\rangle}$};

\node[decision, below=of avgInterval] (checkDs) {$|2C_* - d_s|<10^{-4}$?};

\node[process, left=of checkDs]  (updateDs) {Update $d_s\!\leftarrow\!2C_*$};

\node[process, right=of checkDs] (RG) {Renormalize network in spectral space};

\node[process, below=of RG] (measure) {Measure $d_f$, $d_w$};

\node[decision, below=of measure]    (crossover)  {Crossover in $d_f$, $d_w$?};

\node[process, left=of crossover] (splitTau)
      {Find $\tau_x$, split intervals;\\each interval $\rightarrow$ Init};

\node[decision, below=of crossover] (checkFinal) {$|d_f/d_w - d_s/2|<10^{-1}$?};

\node[terminator, right=of checkFinal] (end) {Accept result};

\node[process, below=of checkFinal] (updateDs2) {Update $d_s\!\leftarrow\! d_f/d_w$ \\ Return to Renormalize network in spectral space};

\draw[arrow] (start) -- (init);
\draw[arrow] (init) -- (selectEig);
\draw[arrow] (selectEig) -- (avgEig);
\draw[arrow] (avgEig) -- (avgInterval);
\draw[arrow] (avgInterval) -- (checkDs);
\draw[arrow] (RG) -- (measure);
\draw[arrow] (measure) -- (crossover);

\draw[arrow] (checkDs.west) -- node[above]{No}  (updateDs.east);
\draw[arrow] (checkDs.east) -- node[above]{Yes} (RG.west);

\draw[arrow] (crossover.west)  -- node[above]{Yes} (splitTau.east);
\draw[arrow] (crossover.south) -- node[right]{No}  (checkFinal.north);

\draw[arrow] (checkFinal.east)  -- node[above]{Yes} (end.west);
\draw[arrow] (checkFinal.south) -- node[right]{No}  (updateDs2.north);

\draw[arrow] (updateDs.west)   -- ++(-5mm,0) |- (selectEig.west);
\draw[arrow] (splitTau.west)   -- ++(-95mm,0) |- (init.west);
\draw[arrow] (updateDs2.east)  -- ++(+50mm,0) |- (RG.east);

\end{tikzpicture}
\end{adjustbox}
\caption{\textbf{Workflow for spectral renormalization and meta-graph reconstruction.} 
It combines $d_s$ estimation, spectral space coarse-graining, crossover detection, 
and consistency validation via critical exponents.}
\label{fig:flowchart_ds_procedure}
\end{figure*}

\clearpage
\newpage




\section{Supernode Distributions for Internet Network}

\subsection{Degree Distribution}

\begin{figure*}[h]
\centering
\includegraphics[width=1.0\linewidth]{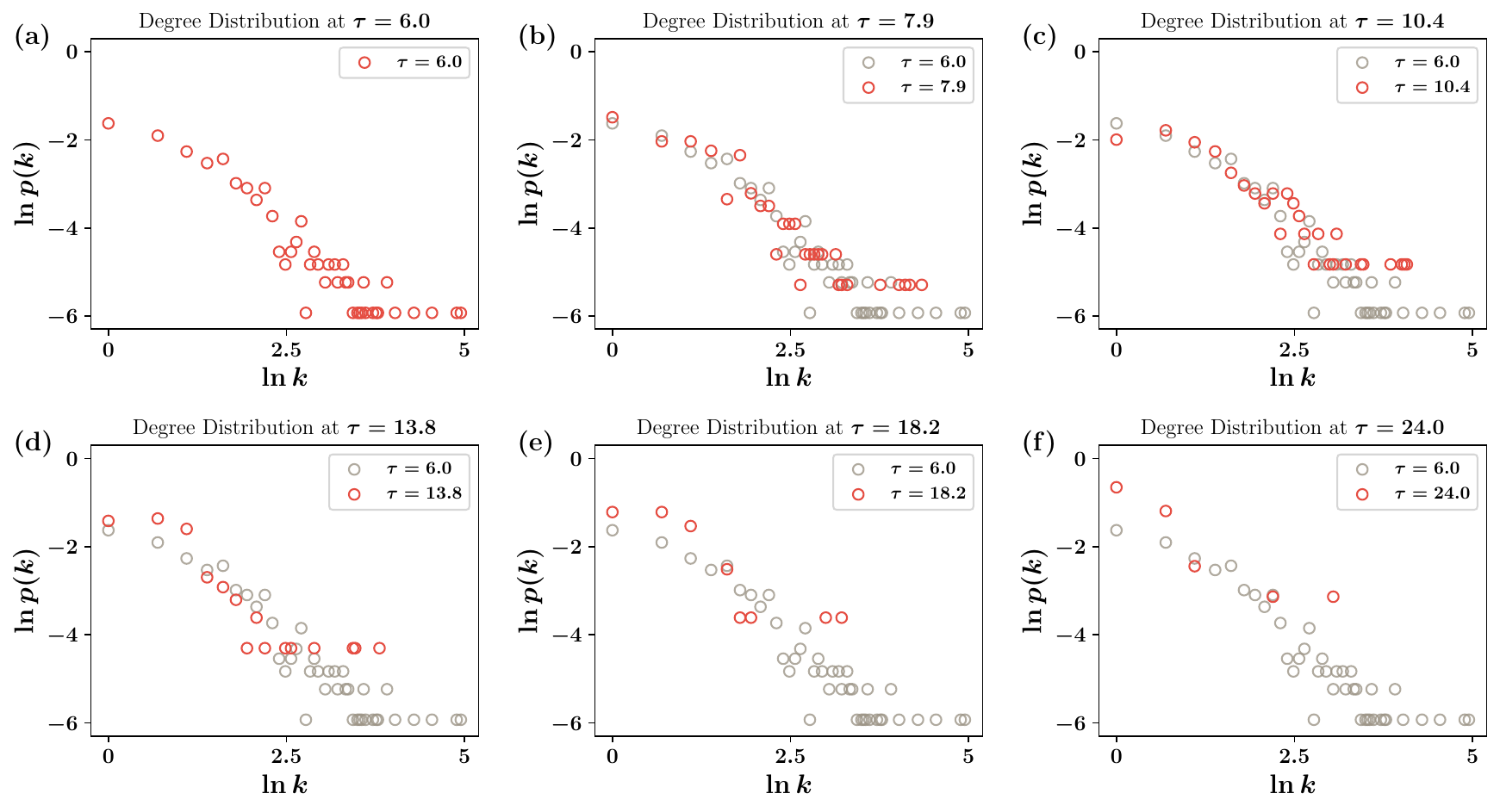}
\caption{\textbf{Degree distributions of supernodes at various $\tau$ in the Internet network (19980101).} 
Each panel shows the log-log plot of $p(k')$ versus $k'$ for a different diffusion time $\tau$, 
revealing how the supernode degree distribution evolves under spectral renormalization. 
The persistence of a heavy-tailed distribution across $\tau$ suggests that the underlying scale-free structure of the network is preserved through the renormalization process.}
\label{fig:internet_degree}
\end{figure*}

\clearpage
\newpage

\subsection{Mass Distribution}

\begin{figure*}[h]
\centering
\includegraphics[width=1.0\linewidth]{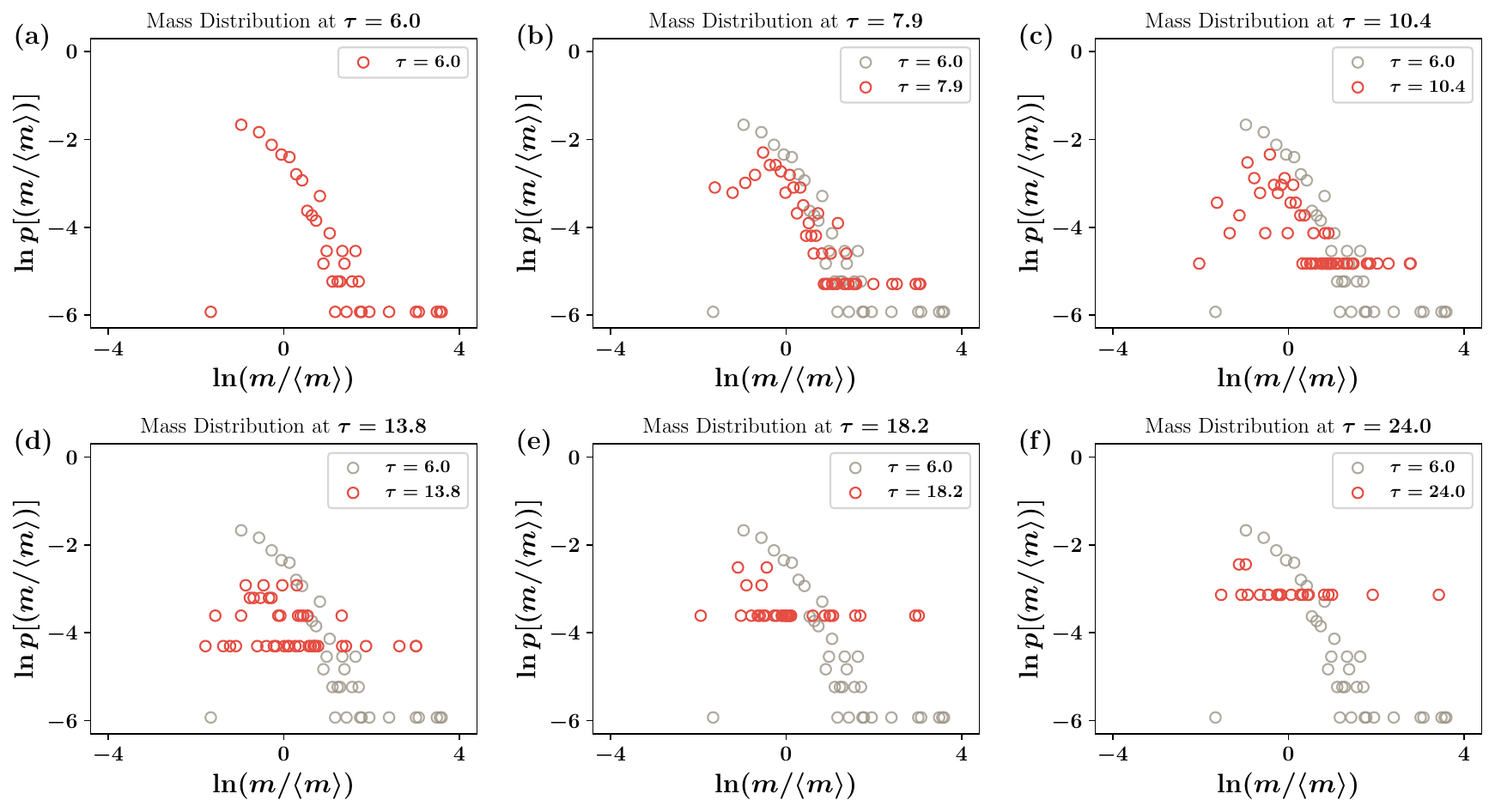}
\caption{\textbf{Normalized mass distributions of supernodes at various $\tau$ in the Internet network (19980101).}
Each panel shows the log-log plot of the distribution $p(m / \langle m \rangle)$ versus normalized supernode mass $m / \langle m \rangle$ at different diffusion times $\tau$. As $\tau$ increases, the mass distribution becomes increasingly heterogeneous, indicating the emergence of dominant supernodes alongside many smaller ones.}
\label{fig:internet_mass}
\end{figure*}

\clearpage
\newpage

\subsection{Size Distribution}

\begin{figure*}[h]
\centering
\includegraphics[width=1.0\linewidth]{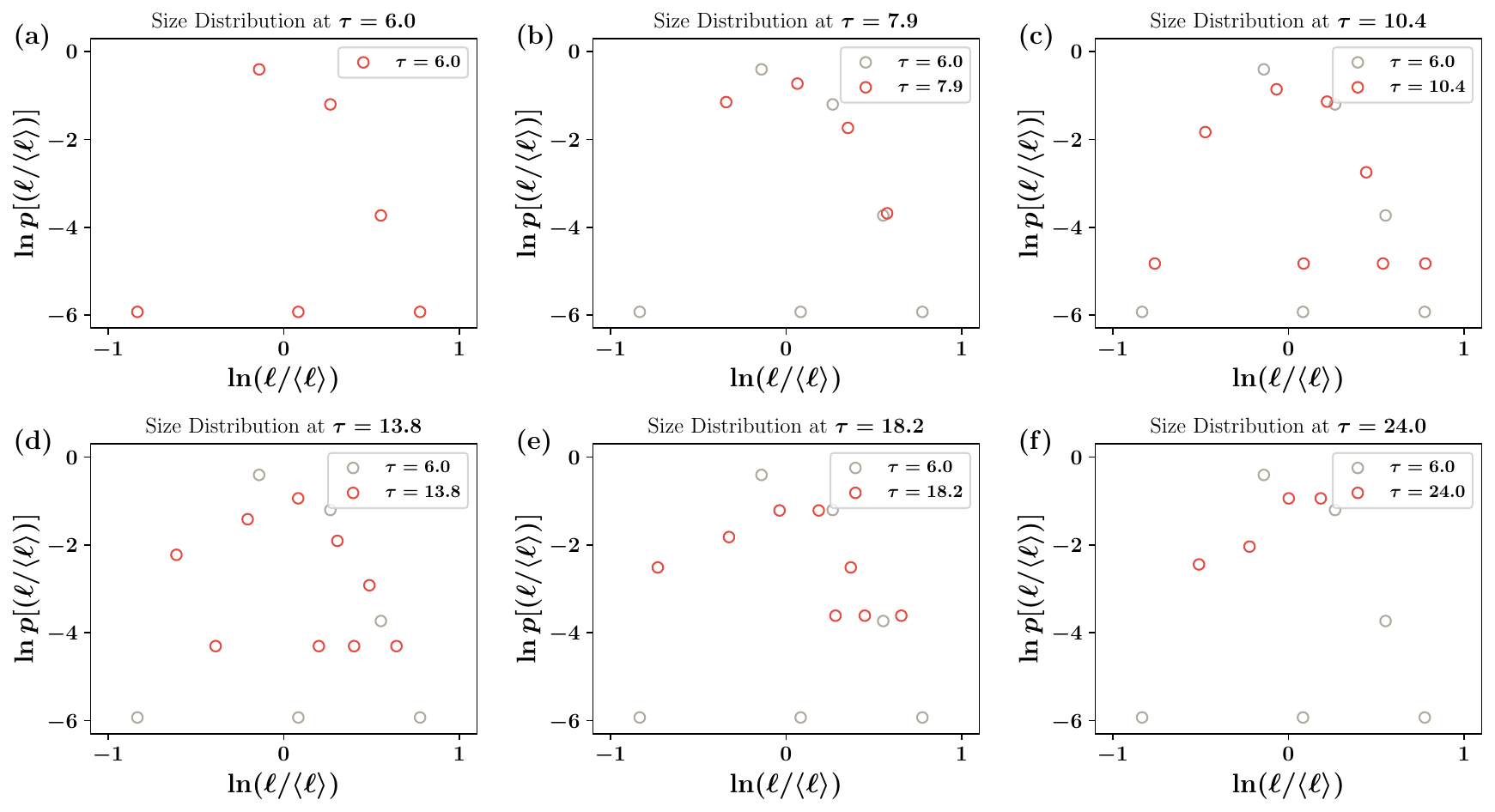}
\caption{\textbf{Normalized size distributions of supernodes at various $\tau$ in the Internet network (19980101).}
Each panel shows the log-log plot of the distribution $p(\ell / \langle \ell \rangle)$ versus normalized supernode size $\ell / \langle \ell \rangle$ at different diffusion times $\tau$.}
\label{fig:internet_size}
\end{figure*}

\clearpage
\newpage


\section{network dimensions and Scaling Relations in Real-World Networks}

\subsection{Yeast Regulatory Network}

\begin{figure*}[h]
\centering
\includegraphics[width=1.0\linewidth]{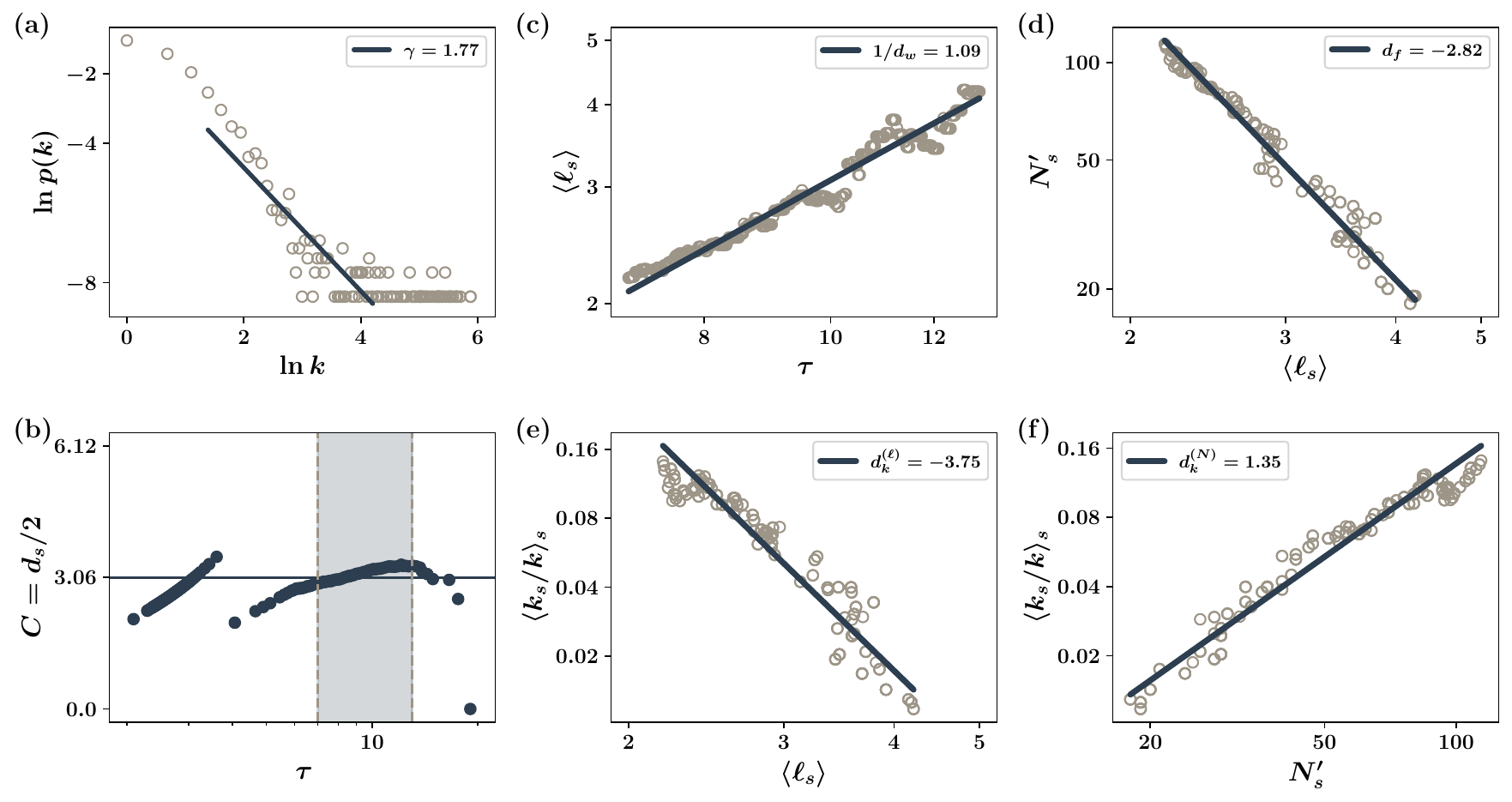}
\caption{
\textbf{Comprehensive scaling analysis extends to biological regulatory networks.} 
Application of our meta-graph algorithm to the yeast transcriptional regulatory network demonstrates broad applicability beyond technological infrastructures.
(a) Biological network exhibits steeper degree distribution with $\gamma_d \approx 1.77$ for $4 \leq k \leq 80$.
(b) Spectral dimension from specific heat plateau: $C = d_s/2 = 3.06$ for $\tau \in [7, 13]$ gives $d_s = 6.12$.
(c) Random-walk dimension: inverse slope $1/d_w = 1.09$ yields $d_w = 0.92$.
(d) Fractal dimension: supernode scaling gives $d_f \approx 2.82$.
(e) Degree scaling consistency: $d_k^{(\ell)} \approx 3.75$.
(f) Theoretical validation: ratio $d_k^{(\ell)}/d_f = 1.33$ closely matches both $d_k^{(N)} = 1.35$ and prediction $1/(\gamma_d - 1) \approx 1.30$.
All independently measured dimensions satisfy the Alexander--Orbach relation ($d_f/d_w = d_s/2$), confirming our comprehensive approach successfully captures scaling behavior across diverse biological and technological network architectures.
}
\label{fig:yeast}
\end{figure*}

\clearpage
\newpage


\subsection{Power grid}

\begin{figure*}[h]
\centering
\includegraphics[width=1.0\linewidth]{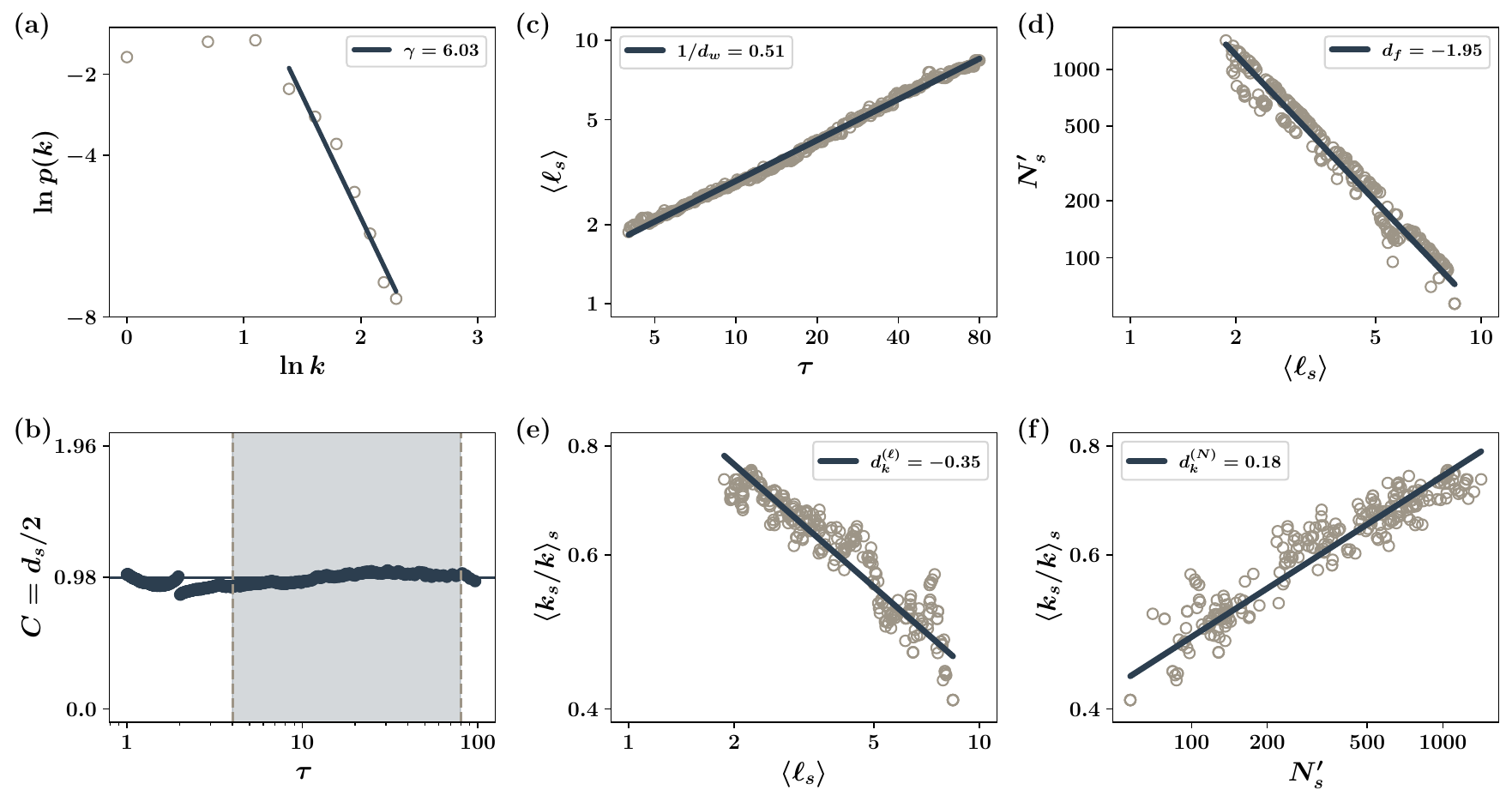}
\caption{
\textbf{comprehensive scaling analysis applies to non-scale-free infrastructure networks.} 
Application of our meta-graph algorithm to the European power grid demonstrates framework robustness beyond power-law degree distributions.
(a) Exponential degree distribution with effective $\gamma_d \approx 6.03$ for $k \geq 4$ differs from typical scale-free form.
(b) Well-defined spectral dimension: plateau at $C = d_s/2 = 0.98$ for extended range $\tau \in [4, 40]$ gives $d_s = 2.00$.
(c) Transport efficiency: random-walk dimension $1/d_w = 0.51$ yields $d_w = 1.96$.
(d) Geometric organization: fractal dimension $d_f \approx 1.95$ indicates near-Euclidean structure.
(e) Degree scaling: $d_k^{(\ell)} \approx 0.35$ reflects infrastructure constraints.
(f) Consistency verification: $d_k^{(\ell)}/d_f = 0.18$ matches both $d_k^{(N)} = 0.18$ and theoretical $1/(\gamma_d - 1) \approx 0.20$.
All independently measured dimensions satisfy the Alexander--Orbach relation, confirming our comprehensive approach successfully captures scaling behavior across scale-free, biological, and infrastructure network architectures.
Additional supernode distributions in Supplementary Figs.~\ref{fig:powergrid_degree}--\ref{fig:powergrid_size}.
}
\label{fig:powergrid}
\end{figure*}

\clearpage
\newpage


\section{Supernode Distributions for Power grid}

\subsection{Degree Distribution}

\begin{figure*}[h]
\centering
\includegraphics[width=1.0\linewidth]{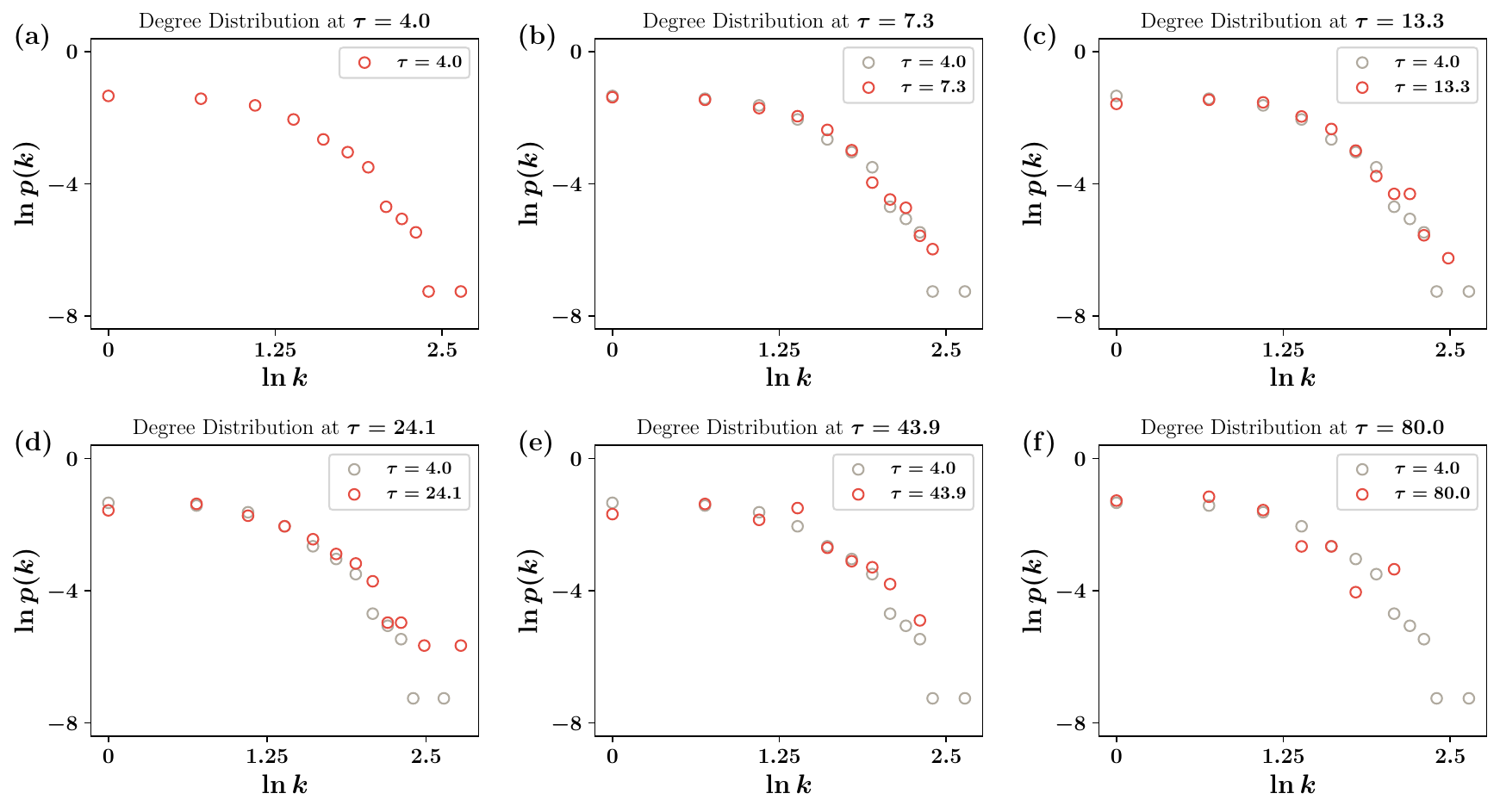}
\caption{\textbf{Degree distributions of supernodes at various $\tau$ in the European power grid.}
Each panel shows the log-log plot of $p(k')$ versus $k'$ for a different diffusion time $\tau$, 
revealing how the supernode degree distribution evolves under spectral renormalization. 
The persistence of a heavy-tailed distribution across $\tau$ suggests that the underlying scale-free structure of the network is preserved through the renormalization process.}
\label{fig:powergrid_degree}
\end{figure*}

\clearpage
\newpage

\subsection{Mass Distribution}

\begin{figure*}[h]
\centering
\includegraphics[width=1.0\linewidth]{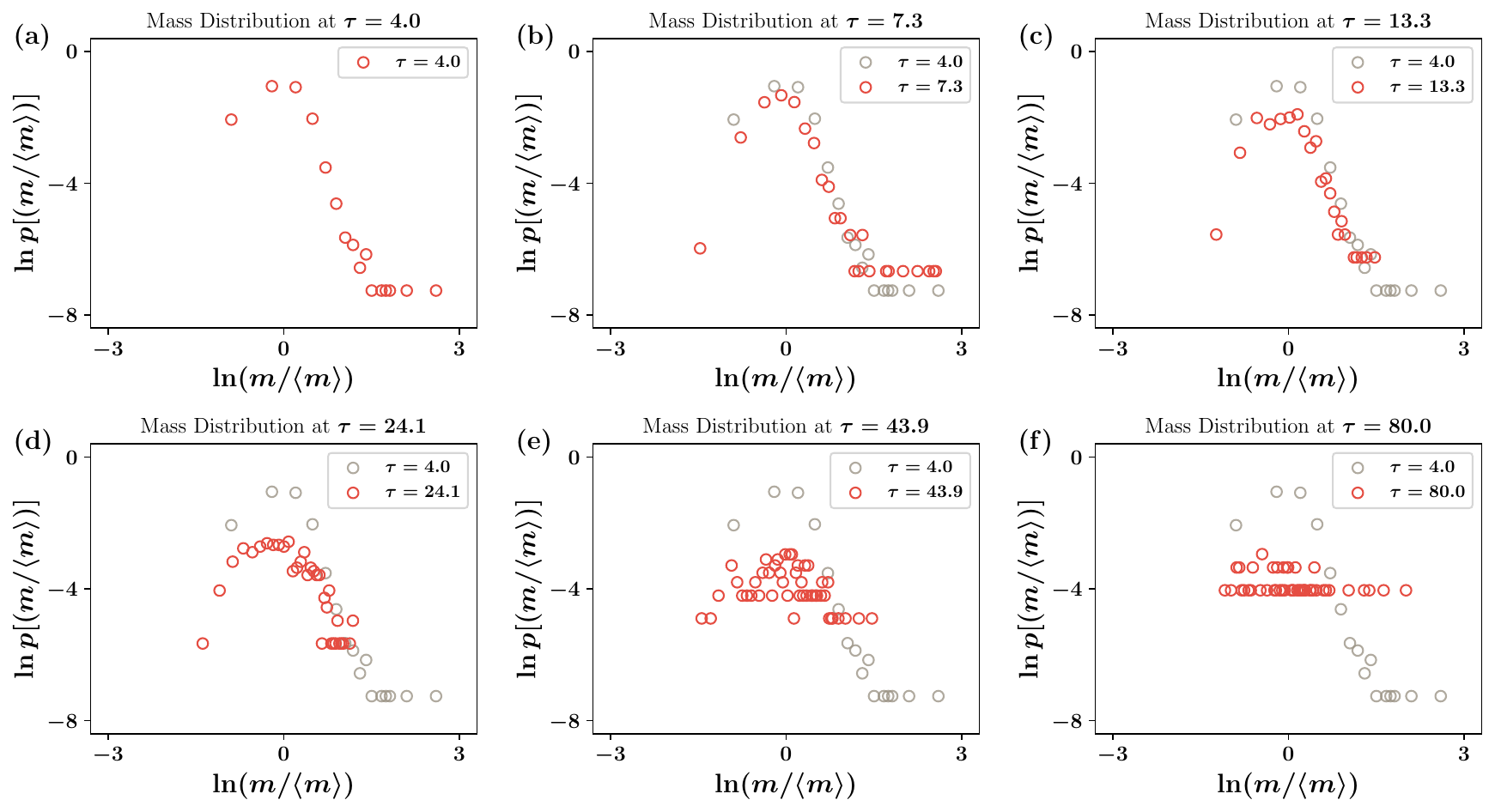}
\caption{\textbf{Normalized mass distributions of supernodes at various $\tau$ in the European power grid.}
Each panel shows the log-log plot of the distribution $p(m / \langle m \rangle)$ versus normalized supernode mass $m / \langle m \rangle$ at different diffusion times $\tau$. As $\tau$ increases, the mass distribution becomes increasingly heterogeneous, indicating the emergence of dominant supernodes alongside many smaller ones.}
\label{fig:powergrid_mass}
\end{figure*}

\clearpage
\newpage

\subsection{Size Distribution}

\begin{figure*}[h]
\centering
\includegraphics[width=1.0\linewidth]{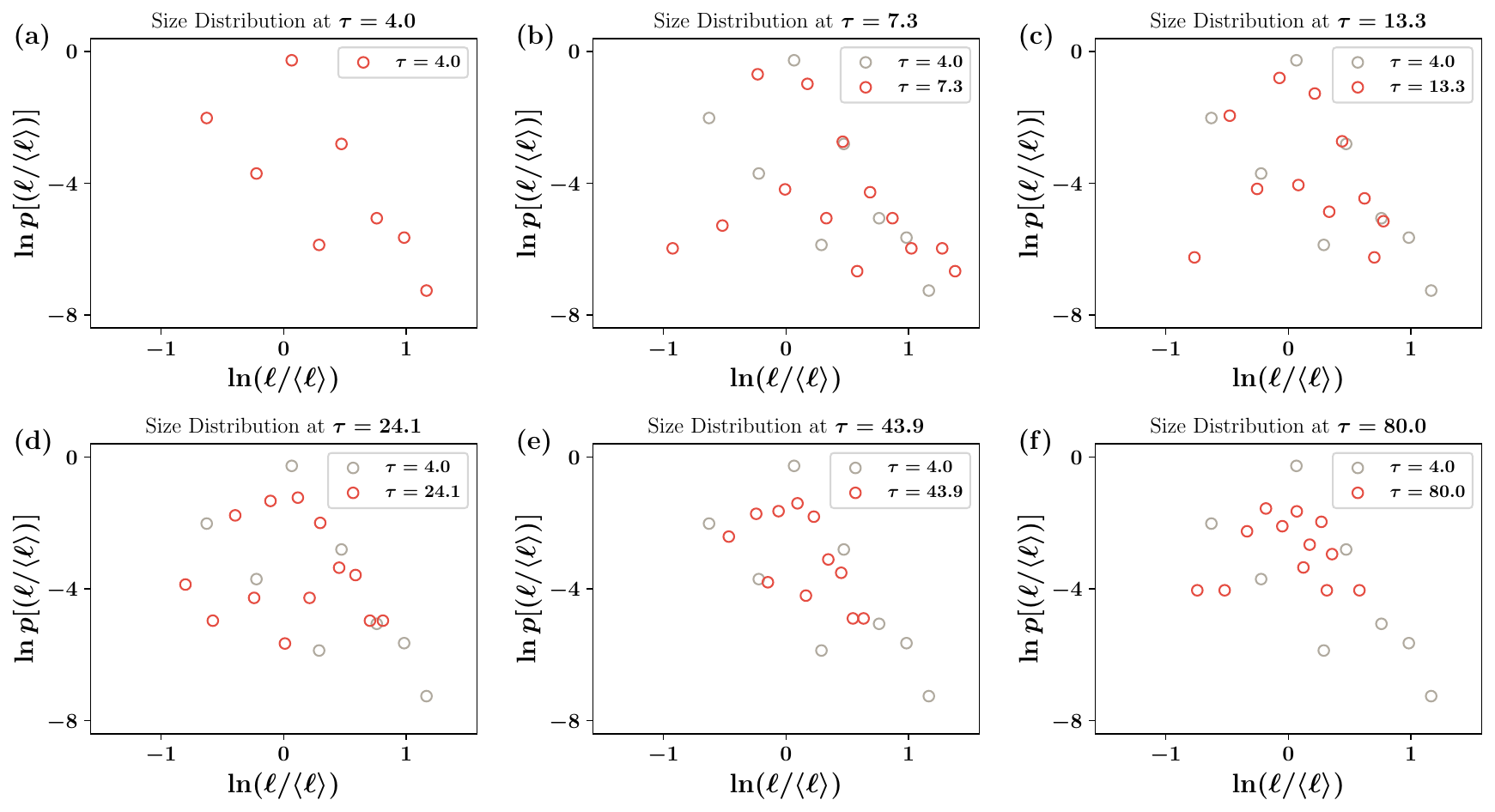}
\caption{\textbf{Normalized size distributions of supernodes at various $\tau$ in the European power grid.}
Each panel shows the log-log plot of the distribution $p(\ell / \langle \ell \rangle)$ versus normalized supernode size $\ell / \langle \ell \rangle$ at different diffusion times $\tau$.}
\label{fig:powergrid_size}
\end{figure*}

\clearpage
\newpage



\section{Multi-Scaling}

\begin{figure*}[h]
\centering
\includegraphics[width=1.0\linewidth]{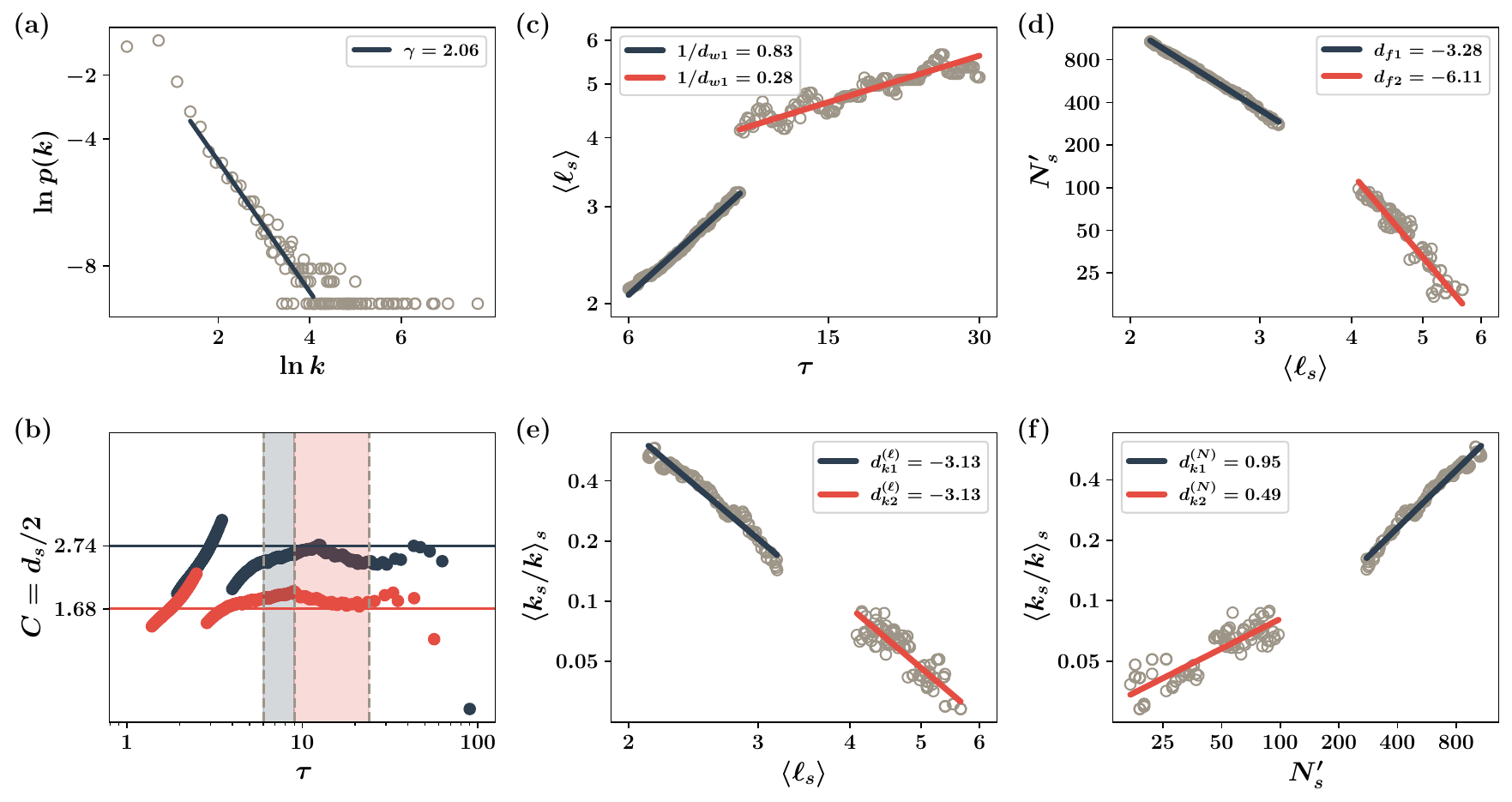}
\caption{
\textbf{Multi-scaling behavior persists throughout Internet evolution.} 
Application of our meta-graph algorithm to Internet AS-level topology (2001) confirms multi-scaling as a robust structural feature rather than a temporal artifact.
(a) Continued evolution toward heavier-tailed distribution: $\gamma_d \approx 2.06$ for $k \in [4, 64]$.
(b) Persistent regime transitions: spectral dimension shifts from $C = d_s/2 = 2.738$ (short $\tau \in [6, 10]$) to $1.680$ (long $\tau \in [10, 30]$).
(c) Transport regime crossover: $1/d_w = 0.83$ vs $0.28$ indicates strengthening of multi-scale structure.
(d) Structural reorganization: $d_f \approx 3.28$ vs $6.11$ shows increasing separation between core and periphery.
(e) Stable degree scaling: $d_k^{(\ell)} \approx 3.13$ maintained across both regimes.
(f) Theoretical consistency: $d_k^{(\ell)}/d_f = 0.95$ matches $d_k^{(N)} = 0.95$ and $1/(\gamma_d - 1) = 0.94$.
The Alexander--Orbach relation remains valid in both regimes despite different exponent values, confirming that multi-scaling represents fundamental structural principles governing Internet evolution rather than measurement artifacts.
}
\label{fig:internet_2001}
\end{figure*}

\clearpage
\newpage


\section{Non-Recursive Nature of the Meta-Graph Algorithm}

We analyze here a property that fundamentally distinguishes the meta-graph algorithm from conventional approaches: its non-recursive nature. This feature explains why iterative procedures break down in spectral space and highlights unique analytical opportunities enabled by our framework.

To illustrate, we examine a deterministic scale-free tree network~\cite{jung2002geometric}, which provides exact self-similarity and thus serves as an ideal testbed. In this model, each node increases its degree by a factor of $m$ at each time step, such that a node created at time $t_a$ has degree $k_i(t) = m^{t - t_a}$ for $t > t_a$. The degree distribution follows a power law, $P(k) \sim k^{-\gamma_d(m)}$, with $\gamma_d(m) = 1 + \ln(2m - 1)/\ln m \to 2$ as $m \to \infty$.

Applying the meta-graph algorithm at varying diffusion times $\tau$ generates distinct renormalized topologies [Fig.~\ref{fig:SF_tree}(c)--(g)]. For small $\tau$, new edges (dashed red lines) appear [panels (c)--(e)] that are absent in the original network [panel (b)]. These edges, induced by spectral correlations, create long-range links that substantially reshape the topology.

Non-recursivity becomes evident in panel (d), where first-generation nodes $\scalebox{1.0}{\textcircled{1}}$ disconnect from the main component. This creates a topological dead-end: configuration (f) cannot be reached from (d) through further iteration. Yet (f) remains directly accessible from the original network [panel (b)] by selecting an appropriate diffusion time $\tau$, as shown in Fig.~\ref{fig:SF_tree}(a). This path dependence—where intermediate renormalized states obstruct access to alternative outcomes—demonstrates that the meta-graph algorithm is fundamentally non-recursive.

This behavior contrasts sharply with conventional real-space RG methods~\cite{kadanoff1966scaling,migdal1996recursion,wilson1975renormalization}, where recursive flows such as (b)$\rightarrow$(f) remain unobstructed. In the meta-graph formulation, global spectral correlations induce irreversible topological deformations, rendering the outcome—and the resulting exponents $d_s$ and $d_f$—highly sensitive to the choice of $\tau$. This underscores the need for scale-parametrized analysis rather than blind iteration.

\begin{figure*}[!h]
\centering
\includegraphics[width=1.0\linewidth]{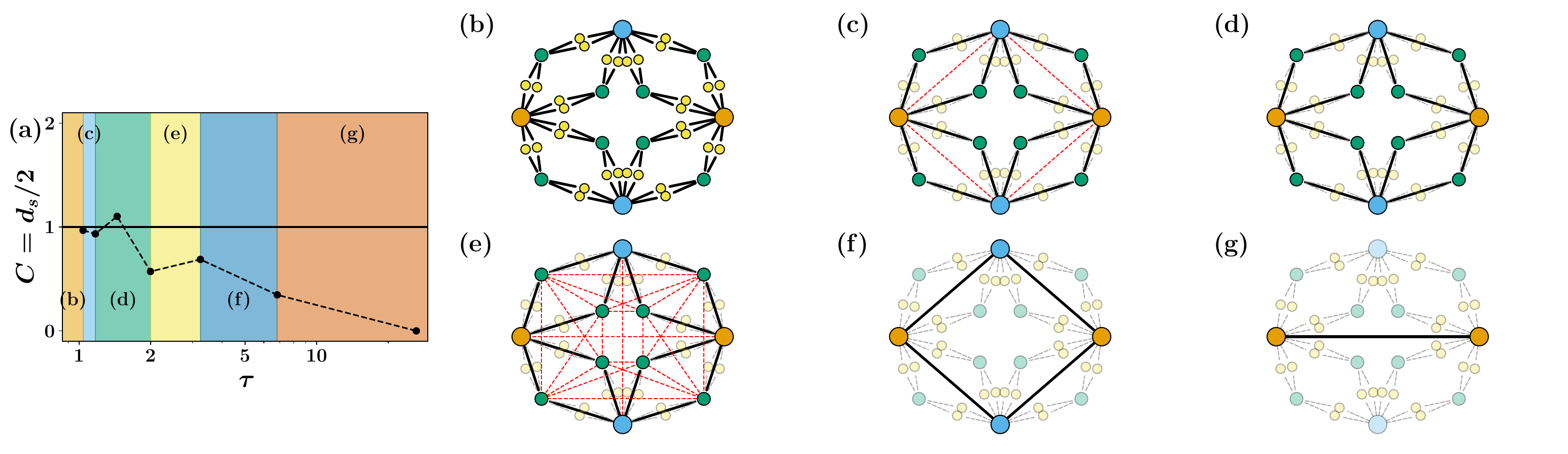}
\caption{
\textbf{Path dependence confirmed in alternative deterministic network architecture.} 
A deterministic scale-free flower network provides independent validation of non-recursive behavior in the meta-graph algorithm.
(a) Time domain mapping shows accessible meta-graph configurations for different diffusion parameters.
(b) Original three-generation flower network with hub-spoke architecture.
(c)--(g) Renormalized topologies demonstrate irreversible structural transformations. Solid lines (thick black) represent preserved connections; dotted lines (thin gray) show removed edges; red dotted lines indicate newly emerged correlations.
Consistent with tree network analysis, specific renormalized configurations block access to other states that remain directly reachable from the original network, confirming path dependence as an intrinsic property of spectral space transformations across different network architectures.
}
\label{fig:SF_loop}
\end{figure*}

A similar analysis of the scale-free flower network (Supplementary Fig.~\ref{fig:SF_loop}) confirms that non-recursivity is a general property of the meta-graph algorithm. Iterative applications inherently suffer from path dependence and irreversibility, making them unsuitable for consistent scaling analysis. By contrast, our scale-parametrized formulation avoids these pitfalls and ensures coherent multiscale behavior.

Although non-recursivity may appear restrictive, it opens analytical opportunities inaccessible to recursive methods. Irreversible deformations induced by spectral correlations expose hidden organizational patterns that local topology alone cannot capture. These patterns underlie the discovery of latent dynamical correlations, as shown in our analysis of critical infrastructure networks in the main text. They reveal dynamically coherent regions with nearly identical spectral signatures despite being spatially distant—reflecting dynamical symmetries rather than static proximity—and offer new insights into both network science and infrastructure resilience.

\clearpage
\newpage



\end{document}